\input epsf
\documentclass[nohyper,notoc]{article} 
\usepackage{axodraw}
\usepackage{epsfig}
\usepackage{bm}
\usepackage{color}

\usepackage[utf8]{inputenc} 
\usepackage[T1]{fontenc}    

\def\ben{\begin{enumerate}}
\def\een{\end{enumerate}}
\def\bit{\begin{itemize}}
\def\eit{\end{itemize}}
\def\beq{\begin{equation}}
\def\eeq{\end{equation}}
\def\bea{\begin{eqnarray}}
\def\eea{\end{eqnarray}}
\def\bq{\begin{quote}}
\def\eq{\end{quote}}
\def \lsim{\mathrel{\vcenter
     {\hbox{$<$}\nointerlineskip\hbox{$\sim$}}}}
\def \gsim{\mathrel{\vcenter
     {\hbox{$>$}\nointerlineskip\hbox{$\sim$}}}}
\def\gappeq{\mathrel{\rlap {\raise.5ex\hbox{$>$}}
{\lower.5ex\hbox{$\sim$}}}}
\def\lappeq{\mathrel{\rlap{\raise.5ex\hbox{$<$}}
{\lower.5ex\hbox{$\sim$}}}}

\def\Zslash{ \, Z  \! \! \! \! / ~ }

\def\meg{\mu \to e \gamma}
\def\teg{\tau \to e \gamma}
\def\tmg{\tau \to \mu \gamma}
\def\Ztm{Z \to \tau^\pm  \mu^\mp }
\def\Zte{Z \to \tau^\pm  e^\mp }
\def\Zme{Z \to \mu^\pm  e^\mp }
\def\tlg{\tau \to \ell \gamma}

\def\meee{\mu \to e \bar{e} e}
\def\tlll{\tau \to 3l }

\def\mec{\mu \to e ~{\rm conversion}}
\def\muc{\mu \to e ~{\rm conversion}}
\def\a{\alpha}
\def\b{\beta}
\def\g{\gamma}

\def\m{\mu}

\begin{document}

\renewcommand{\thefootnote}{\fnsymbol{footnote}}
\begin{center}
{\Large  {\bf
$\bm{\mu \to e \gamma} $  and matching at $\bm{m_W}$ 
}}

\vskip 25pt
{\bf   Sacha Davidson \footnote{E-mail address:
s.davidson@ipnl.in2p3.fr}  
}
\vskip 10pt  
{\it IPNL, CNRS/IN2P3,  4 rue E. Fermi, 69622 Villeurbanne cedex, France; 
Universit\'e Lyon 1, Villeurbanne;
 Universit\'e de Lyon, F-69622, Lyon, France
}\\
\vskip 20pt
\vskip 20pt
{\bf Abstract}
\end{center}

\begin{quotation}
  {\noindent\small 
Several experiments search for $\mu \leftrightarrow e$ flavour change,
for instance  in $\mec$, $\meg$ and $\meee$. 
This paper studies how to translate these  experimental 
constraints from low energy to a New Physics scale $M \gg m_W$. 
A basis of QCD$\times$QED-invariant operators (as appropriate below $m_W$)
is reviewed,  then run to $m_W$ with
one-loop Renormalisation Group Equations (RGEs)
 of QCD and QED.  At $m_W$,  these operators
are  matched onto  SU(2)-invariant dimension-six operators, 
which can continue to run up with electroweak RGEs.  
As an example,
the $\meg$ bound is translated  to  the scale $M$,  where it constrains 
two sums of operators. 
The constraints differ from  those obtained in previous EFT analyses of $\meg$,
but reproduce the expected bounds on
flavour-changing interactions of  the $Z$ and the Higgs,
because the matching at $m_W$ is pragmatically performed
to the loop order required to get the ``leading'' contribution.

\vskip 10pt
\noindent
}

\end{quotation}

\section{Introduction}
\label{intro}


Neutrino masses and mixing angles imply that ``New'' Physics 
from beyond the Standard Model(SM) must be present in the lepton sector,
and must induce charged  Lepton Flavour Violation(LFV;
for a review, see \cite{KO}). 
However, neither LFV nor the origin of neutrino
masses has yet been discovered. 
This study assumes that the required new particles  
are heavy, with 
masses at or beyond  $M > m_W$. In addition, 
between $m_W$ and $M$, there
should be no other  new particles or interactions
 which affect the
LFV sector. 
One approach to identifying this New LFV  Physics, is
to construct  a motivated model, and identify
its signature  in  observables
\cite{composite,SUSY,BR,vector,RS,inverse,seesaw,higgs,4gen,ZeeBabu}.
A more pragmatic approach, which  requires optimism but no
model-building skills, is to parametrise the
New Physics at low energy  with non-renormalisable
operators, map the experimental constraints
onto the operator coefficients,  and   attempt to reconstruct
the fundamental  Lagrangian of New Physics from
the operator coefficients. This
is probably not feasible, but could give interesting
perspectives.  A first step
in this ``bottom-up'' approach, explored
in this paper,  is to  use Effective Field Theory (EFT)
\cite{Georgi} to translate the
experimental   bounds to the  coefficients of
effective operators  at the 
New Physics  scale $M> m_W$.

The  goal  would be   to start  from experimental constraints 
on $\mu-e$ flavour change,
and obtain   at  $M$  the best  bound on each
coefficient from each observable.
These constraints should be of the correct
order of magnitude, 
but not precise beyond
one significant figure.   This preliminary study
restricts  the experimental input to the
bound  on $BR(\meg)$, and makes several simplifications
in the translation up to the 
New Physics  scale $M$.
Firstly, the EFT has  three scales: a low scale $m_\mu \sim m_b$,
the intermediate weak scale $m_W$, and the high scale  $M$. 
Secondly, at a given scale,
the EFT contains lighter Standard Model particles
and   dimension six, gauge-invariant operators
(one dimension seven operator
is listed; however 
dimension eight operators are  neglected).
The final simplification  might have been  to match at tree-level, 
and run with one-loop Renormalisation Group Equations (RGEs).
However, a  bottom-up EFT should reproduce the
results of top-down  model calculations,
and it is straightforward to check that
 one  and  two-loop matching  is
required at $m_W$ to  obtain the
correct bounds from $\meg$ on 
LFV interactions of the $Z$ and Higgs.
So the matching at $m_W$ is
performed to the order required to 
get the known bounds.

The paper is organised in two parts: the
first sections \ref{sec:low} - \ref{@mW} construct  some of the framework 
required to  obtain experimental constraints
on SU(2) invariant operator coefficients at $m_W$,
then  section \ref{sec:meg} focusses on
using, checking and improving this formalism
to obtain bounds from $\meg$ on operator
coefficients at $M$.
The formalism can be organised
in four steps:
 matching at $m_\mu$, running to $m_W$, 
matching at $m_W$,  then running  up to  the New
Physics scale $M$.
Section \ref{sec:low} reviews 
 the  basis of QCD$\times$QED invariant operators,
as appropriate below $m_W$. These operators, of
dimension five, six and seven,
describe three and four-point functions
involving a $\mu$, an $e$ and any  other
combination  of flavour-diagonal  light
particles. 
To complete the first step,  the  experimental bounds should be
matched onto these operator coefficients;
however, this is delayed til  section  \ref{sec:meg},
where  only the bound
on $\meg$ is imposed on the dipole coefficients
(the bounds from $\mec$ and $\meee$ are neglected
for simplicity; the strong interaction subtleties of
matching to $\muc$ are discussed in 
\cite{KKO,CKOT,CrivellinMuc}).
Section \ref{sec:runQED} discusses
the second  step, which is to run the coefficients up to $m_W$
with the RGEs of QED and QCD. Appendix \ref{sec:anomdim}
gives the anomalous dimension matrix 
mixing the  scalar and tensor operators
to the dipole (which is responsable for $\meg$). 
The anomalous dimension matrix for  vector operators is neglected
for two reasons: although vectors contribute at tree level to
$\mec$ and $\meee$,  these experimental
bounds are not included, and the leading order mixing of vectors
 to the dipole
is at two-loop in QED, whereas the running here
is only performed at one-loop.
The next step is to match  these operators
at $m_W$  onto the  Buchmuller-Wyler\cite{BW} 
basis of SU(2) invariant operators
as pruned in \cite{polonais}, which is
refered to as the BWP basis.   The tree-level matching
for all operators is given in section 
 \ref{@mW};   if this is
the leading contribution
to the coefficients, then imposing SU(2)
invariance above $m_W$ predicts some
ratios of coefficients below $m_W$,
as discusssed in section \ref{ssec:comments}.
Section  \ref{sec:meg} 
uses the formalism of the
previous sections to
 translates the
experimental bound on $BR(\meg)$
to sums of  SU(2)-invariant operator coefficients at $m_W$.
Then  a few finite loop
contributions are added, and 
the  coefficients are
run up  to $M$, using a simplified version
of the one loop QCD and electroweak  RGEs \cite{PS,JMT}.
 Finally, section \ref{sec:disc}
discusses various  questions arising from
this study, such as the  loop order
required in matching at $m_W$,
whether the  non-SU(2)-invariant
basis is required below $m_W$, and  the
importance of QED running  below $m_W$.

Many parts  or this analysis can be found  in previous  literature. 
Czarnecki and Jankowski\cite{CzarJ} emphasized
the one-loop  QED running of the dipole operator
(neglected in the estimates here), which 
shrinks the coefficient at low energy.
Degrassi and Giudice\cite{PepeGian} give the leading order
QED mixing of vector operators to the dipole,
which is also neglected here, because it
arises at two-loop.
In an early top-down  analysis, Brignole and
Rossi \cite{BR} calculated a  wide
variety of LFV processes   as a function
of  operator coefficients above
$m_W$, without explicit Renormalisation
Group running and a slightly redundant
basis.  Pruna and Signer \cite{PS} studied
$\meg$ in EFT,    focussing on the electroweak running
above  $m_W$, which they
perform in more detail than is done here. 
However, they do not obtain  the bounds on
the  LFV couplings of the $Z$  and Higgs 
that arise here in matching at $m_W$.
 Various one-loop contributions to $\meg$
were calculated in \cite{CNR}, without organising them  into
running and matching parts.
Finally, the contribution of the
 LFV Higgs operator to LFV $Z$ couplings
was beautifully studied in \cite{GKM}.
 There are also many closely related works in
the quark sector, reviewed in \cite{BBL,BurasHouches}.
For instance, the QED anomalous dimension matrix
for various vector four-quark operators 
is given in \cite{Lusignoli}, 
and matching at $m_W$  of
flavour-changing quark operators
is discussed in \cite{ACFG}.
However, colour makes the quarks
 different, so it is not 
always immediate to translate the
quark results to leptons.


\section{A basis of   $\mu-e$ interactions  at low energy }
\label{sec:low}

\subsection{Interactions probed in muon experiments }
\label{ssec:belowmtau}

Experiments searching for lepton flavour change  from  $\mu$ to $e$, 
probe three- and
four-point functions involving a muon, an electron and
one or two other SM particles.  I focus here on interactions
that can be probed in $\meg$, $\meee$ and $\muc$,
meaning that the interactions  are otherwise flavour
diagonal, and there is only one muon (so $K\to \bar{\mu}e$
and other meson decays is not considered).

These ``new physics''  interactions
can be written  as non-renormalisable operators
involving  a  single $\mu$, and  some combination of 
$e,\g,g, u,d,$ or $s$.  The operators 
should   be QED and QCD invariant
(because  we are intested in LFV, not departures from the
SM gauge symmetries), and can be
of any dimension (because the aim is to
list the three-point and four-point interactions
that the data constrains). 
The list, which  can be found in \cite{KO,KKO,CKOT} but
with different names, is:

\beq
\begin{array}{llll}
{\rm dipole} &{\cal O}^{e\mu}_{D,Y } =  {\color{black}  m_\mu } (\overline{e}  \sigma^{\a \b}P_Y  \mu ) F_{\a \b}
\nonumber \\
&&& \nonumber \\
{\rm 4~lepton}
 &{\cal O}^{e\mu ee}_{YY} = \frac{1}{2}(\overline{e} \gamma^\a P_Y \mu ) 
(\overline{e} \gamma^\a P_Y e )  &,&
{\cal O}^{e\mu ee}_{YX} = \frac{1}{2}(\overline{e} \gamma^\a P_Y \mu ) 
(\overline{e} \gamma^\a P_X e )  \nonumber \\
&{\cal O}^{e\mu ee}_{S,YY} = (\overline{e}  P_Y \mu ) 
(\overline{e} P_Y e ) &&\\
&&& \nonumber \\
{\rm 2~lepton~2~quark}
&{\cal O}^{e\mu uu}_{YY} = \frac{1}{2}(\overline{e} \gamma^\a P_Y \mu ) 
(\overline{u} \gamma^\a P_Y u )  &,&
{\cal O}^{e\mu uu}_{YX} = \frac{1}{2}(\overline{e} \gamma^\a P_Y \mu ) 
(\overline{u} \gamma^\a P_X u )  \nonumber \\
&{\cal O}^{e\mu uu}_{S,YY} = (\overline{e}  P_Y \mu ) 
(\overline{u} P_Y u ) &,&  
{\cal O}^{e\mu uu}_{S,YX} = (\overline{e}  P_Y \mu ) 
(\overline{u} P_X u )  \nonumber \\
&{\cal O}^{e\mu uu}_{T,YY} = (\overline{e} \sigma P_Y \mu ) 
(\overline{u} \sigma P_Y u ) && \nonumber \\
&{\cal O}^{e\mu dd}_{YY} = \frac{1}{2}(\overline{e} \gamma^\a P_Y \mu ) 
(\overline{d} \gamma^\a P_Y d )  &&
{\cal O}^{e\mu dd}_{YX} = \frac{1}{2}(\overline{e} \gamma^\a P_Y \mu ) 
(\overline{d} \gamma^\a P_X d )  \nonumber \\
&{\cal O}^{e\mu dd}_{S,YY} = (\overline{e}  P_Y \mu ) 
(\overline{d} P_Y d )  &&
{\cal O}^{e\mu dd}_{S,YX} = (\overline{e}  P_Y \mu ) 
(\overline{d} P_X d )  \nonumber \\
&{\cal O}^{e\mu dd}_{T,YY} = (\overline{e} \sigma P_Y \mu ) 
(\overline{d} \sigma P_Y d ) &&   \nonumber \\
&{\cal O}^{e\mu ss}_{YY} = \frac{1}{2}(\overline{e} \gamma^\a P_Y \mu ) 
(\overline{s} \gamma^\a P_Y s )  &&
{\cal O}^{e\mu ss}_{YX} = \frac{1}{2}(\overline{e} \gamma^\a P_Y \mu ) 
(\overline{s} \gamma^\a P_X s )  \nonumber \\
&{\cal O}^{e\mu ss}_{S,YY} = (\overline{e}  P_Y \mu ) 
(\overline{s} P_Y s )  &&
{\cal O}^{e\mu ss}_{S,YX} = (\overline{e}  P_Y \mu ) 
(\overline{s} P_X s )  \nonumber \\
&{\cal O}^{e\mu ss}_{T,YY} = (\overline{e} \sigma P_Y \mu ) 
(\overline{s} \sigma P_Y s ) &&   \nonumber \\
&&&   \nonumber \\
{\rm 2~lepton~2~boson}& 
{\cal O}^{e\mu}_{GG,Y} = \frac{1}{M}(\overline{e}  P_Y \mu ) G_{\a \b} G^{\a\b} 
&,& {\cal O}^{e\mu}_{G\tilde{G},Y} = \frac{1}{M}
(\overline{e}  P_Y \mu ) G_{\a \b} \widetilde{G}^{\a\b} 
\nonumber
\end{array}
\label{obsops}
\eeq
where $X,Y \in \{L,R\}$, and $X\neq Y$. 
 These operators are chosen,
using Fiertz and other spinor identities,  to
always have the lepton flavour-change inside a spinor contraction.
Notice also that, following Kuno and Okada \cite{KO}, the
dipole is normalised with a muon Yukawa coupling.
The four-fermion 
operators are labelled with the fermion flavours in superscript,
and in the
subscript is the type of Lorentz contraction
(Scalar, Tensor or Vector --- except the vector
case is implicit), followed
by the chiralities of the two fermion contractions in subscript.
The Lorentz contraction --- Dipole, Scalar, Tensor
or vector---will be used through this paper to categorise
operators.
The operator coefficients have the same
index structure, so $C^{ijkl}_{XX}$ is the coefficient
of  ${\cal O}^{ijkl}_{XX}$, which is a vector contraction
of fermions of chirality $X$.

All the operators appear in the Lagrangian with a 
coefficient $-C/M^2$, and the operator normalisation is chosen
to ensure that the Feynman rule is $-i C/M^2$. This implies
a judicious distribution of $\frac{1}{2}$s, which
is discussed in Appendix \ref{app:2}.

Obtaining constraints from data on the operator
coefficients is reviewed in \cite{KO}, and
$\muc$ is discussed in \cite{KKO,CKOT}. 
Searches for  $\meg$ probe the dipole operator,
 $\meee$ probes the four-lepton operators
and the (off-shell)  dipole, and
$\muc$ probes the two-quark-two-lepton, diboson
and  dipole operators. 
It is interesting that these three
processes are sensitive to almost
all the three-and four-point functions
that involve one muon, any of
the lighter fermions, or photons or gluons.
The only three- or four-point interactions not probed at tree level are
the two-photon interactions
$$
 {\cal O}^{e\mu}_{FF,Y} =
(\overline{e}  P_Y \mu ) F_{\a \b} F^{\a\b} 
~~~,~~~ {\cal O}^{e\mu}_{F\tilde{F},Y} = 
(\overline{e}  P_Y \mu ) F_{\a \b} \widetilde{F}^{\a\b} 
~~~.
$$

\subsection{Including heavy fermions}
\label{@mtau}

At a slightly higher scale, operators containing
$c,b$ $\mu$ and $\tau$ bilinears should be
included.  
These additional operators are:
\beq
\begin{array}{llll}
{\rm 4~lepton}
 &{\cal O}^{e\mu ll}_{YY} = \frac{1}{2}(\overline{e} \gamma^\a P_Y \mu ) 
(\overline{l} \gamma^\a P_Y l )  &,&
{\cal O}^{e\mu ll}_{YX} = \frac{1}{2}(\overline{e} \gamma^\a P_Y \mu ) 
(\overline{l} \gamma^\a P_X l )  \nonumber \\
&{\cal O}^{e\mu ll}_{S,YY} = (\overline{e}  P_Y \mu ) 
(\overline{l} P_Y l ) &&
{\cal O}^{e\mu \tau \tau}_{S,YX} = (\overline{e}  P_Y \mu ) 
(\overline{\tau} P_X \tau )
 \nonumber \\
&{\cal O}^{e\mu \tau \tau}_{T,YY} = (\overline{e} \sigma P_Y \mu ) 
(\overline{\tau} \sigma P_Y \tau )&& \nonumber \\
&&& \nonumber \\
{\rm 2~lepton~2~quark}
&{\cal O}^{e\mu qq}_{YY} = \frac{1}{2}(\overline{e} \gamma^\a P_Y \mu ) 
(\overline{q} \gamma^\a P_Y q )  &,&
{\cal O}^{e\mu qq}_{YX} = \frac{1}{2}(\overline{e} \gamma^\a P_Y \mu ) 
(\overline{q} \gamma^\a P_X q )  \nonumber \\
&{\cal O}^{e\mu qq}_{S,YY} = (\overline{e}  P_Y \mu ) 
(\overline{q} P_Y q ) &,&  
{\cal O}^{e\mu qq}_{S,YX} = (\overline{e}  P_Y \mu ) 
(\overline{q} P_X q )  \nonumber \\
&{\cal O}^{e\mu qq}_{T,YY} = (\overline{e} \sigma P_Y \mu ) 
(\overline{q} \sigma P_Y q ) && 
\end{array}
\label{opstauW}
\eeq
where $l \in \{\mu,\tau\}$,
$q \in \{c,b\}$, $X,Y \in \{L,R\}$, and $X\neq Y$.

Including these operators introduces a second ``low'' scale into
the EFT, which in principle changes the running and
requires matching at this second low  scale $m_\tau$.
The running is discussed in the next section.
Since the matching is at tree-level,  
the operators
present below $m_\tau$ have  the same coefficient
just above $m_\tau$.  
Were the dipole to be matched  at one loop, 
then  at $m_\tau$,  one should 
compute the finite part of the diagrams\cite{CNR} obtained by
closing 
the heavy fermion loop of the tensors ${\cal O}^{e\mu bb}_{T,YY}$,
 ${\cal O}^{e\mu cc}_{T,YY}$ and ${\cal O}^{e\mu \tau \tau }_{T,YY}$,
and attaching a photon (and also there could
be similar finite  contributions from
four-fermion operators at $m_\mu$).
Also,  scalar operators involving $b,c$ quarks
would match at one loop  onto ${\cal O}^{GG,Y}$
\cite{SVZ78}, as outlined in\cite{HisanoDM}.


\section{ Running up to   $ m_W$ }
\label{sec:runQED}

The operators of eqns 
(\ref{obsops}),(\ref{opstauW}) can evolve with scale due
to QED and QCD interactions.  QCD effects
can be significant, and should be resummed,
but fortunately they only change
the magnitude of operator coefficients,
without mixing one operator into another.
This will be taken into account by multiplying
two-lepton-two-quark operators
  by an appropriate factor
(following Cirigliano {\it et.al.} \cite{CKOT}).
The effects of QED running are usually small,
of order $\alpha_{em}/\pi$,  but interesting
because they give operator mixing. Therefore
the QED renormalisation of individual
operator coefficients is neglected, and
only the mixing is included.

The scale at which the operators of eqns 
(\ref{obsops}),(\ref{opstauW})  start running is variable.
The lepton operators of eqn (\ref{obsops}) will start
their QED running at $m_\mu$, whereas those of
eqn (\ref{opstauW}) start at $m_\tau$.  The
the two-lepton-two-$b$ operators start
running up at $m_b$. For simplicity,
the remaining two-quark-two-lepton operators
are taken to start running up  at $m_\tau$;
that is, the experimental bounds are assumed
to apply at a scale $\sim m_\tau$.

\subsection{Defining the  anomalous dimension matrix}

After including one-loop corrections in $\overline{MS}$,  the operator
coefficients will run with scale $\mu$ according to
\beq
\mu \frac{\partial}{\partial \mu} \vec{C} =  \frac{\alpha_{s}}{4\pi}\vec{C} \bm{\Gamma}^{s} +  \frac{\alpha_{em}}{4\pi}\vec{C} \bm{\Gamma} 
\label{RGE}
\eeq 
where the  coefficients of all the
operators listed in the previous section have  been organised into a row vector
$ \vec{C} $,
and 
 $\frac{\alpha_{em}}{4\pi} \bm{\Gamma}$ is
the anomalous dimension matrix, which is
calculated 
as discussed in \cite{BurasHouches}\footnote{
Generically, the one-loop corrections to an 
operator $Q$ will generate divergent coefficients
for  other operators $\{ B\}$.
If one computes the one-loop corrections to  the amputed  Greens function
for the  operator $Q$, with $n$ external legs,
 and Feynman rule $i f_Q Q$, these can be written as
$ i f_Q  \frac{\alpha}{4\pi} \frac{1}{\epsilon}
\sum_B b_{QB} B$. Then $[\bm{\Gamma}]_{QB} =  
{\color{black} -} 2 [ b_{QB} +  \frac{n}{2} a \delta_{QB}] $
where $-\frac{\alpha}{4\pi} a = \frac{\mu}{Z} \frac{\partial} {\partial \mu} 
Z$, and  $Z$  renormalises the   wave-function.}.

The eqn (\ref{RGE}) can be approximately solved,
by neglecting the scale-dependance of $\alpha_{em}$ 
and defining the eigenvalues of the diagonal
 $\bm{\Gamma}^s$  to be
$\{\gamma^s_A\}$,  as:
\beq
   C_A(m_W)\left[\frac{ \alpha_s(m_W)}{\alpha_s(m_\tau)}\right]
^{{\color{black} - \frac{\gamma^s_A}{2 \beta_0}}}
\left(
\delta_{AB} - \frac{\alpha_{em}  }{4\pi} \left[ \Gamma\right]_{AB}
\log \frac{m_W}{m_\tau}
 + 
\frac{\alpha_{em} ^2 }{32\pi^2} \left[ \Gamma \Gamma\right]_{AB}
\log ^2  \frac{m_W}{m_\tau} +..
\right)
= C_B (m_\tau)
\label{oprun1l}
\eeq
where $\beta_0 = 11- 2N_f/3$ from the QCD $\beta$-function,
and $\log   \frac{m_W}{m_\tau} \simeq 3.85$.

It is convenient to separate
the vector of coefficients below  $m_W$,
 $ \vec{C}(<m_W)$, into
subvectors:
\bea
\vec{C}(<m_W) &=& (\vec{C}^{q_i}_V, \vec{C}^{u}_V,  \vec{C}^{d}_V,  \vec{C}^{\tau}_V,  \vec{C}^{\mu}_V, \vec{C}^{e}_V,
\vec{C}^{q_i}_S, \vec{C}^{u}_S,  \vec{C}^{d}_S,  \vec{C}^{\tau}_S,  \vec{C}^{\mu}_{S \mu}, \vec{C}^{e}_{S e},
\vec{C}^{q_i}_T, \vec{C}^{u}_T,  \vec{C}^{d}_T,  \vec{C}^{\tau}_T, \vec{C}_D,
\vec{C}_{GG})
\label{longcoefficientvector} \\    
\vec{C}_V^{ f}& = &( C_{LL}^{e \mu f f} , C_{RR}^{e \mu f f},
C_{LR}^{e \mu f f}, C_{RL}^{e \mu f f}) \\
\vec{C}_S^{f}& = &( C_{S,LL}^{e \mu f f} , C_{S,RR}^{e \mu f f},
C_{S,LR}^{e \mu f f}, C_{S,RL}^{e \mu f f}) ~~~{\rm for}~
f \in \{ q_i,u,d,\tau \} \\
\vec{C}_{Sl}^l& = &( C_{S,LL}^{e \mu ll} , C_{S,RR}^{e \mu ll})
~~~{\rm for}~
l \in \{ e,\mu \} \\
\vec{C}_T^{f} & = &( C_{T,LL}^{e \mu ff}  , C_{T,RR}^{e \mu ff} ) \\
\vec{C}_D & = &( C^{e \mu } _{D,L} , C^{e \mu } _{D,R}) \\
\vec{C}_{GG} & = &( C^{e \mu }_{GG,L} , C^{e \mu }_{GG,R}
)
\eea

\begin{figure}[h]
\begin{center}
\begin{picture}(70,70)
\GCirc(20,47){5}{.7}
\ArrowLine(0,60)(20,50)
\ArrowLine(20,50)(40,60)
\put(-6,60){$\mu$} 
\put(45,60){$e$} 
\ArrowLine(0,-10)(20,0) 
\ArrowLine(20,0)(40,-10)
\ArrowArc(20,30)(15,0,360)
\put(35,31){$f_1$} 
\put(-16,-12){$f_2$} 
\put(46,-10){$f_2$} 
\Photon(20,15)(20,0){2}{3} 
\end{picture}
\hspace{2cm}
\begin{picture}(70,70)
\GCirc(20,47){5}{.7}
\ArrowLine(0,60)(17,47)
\ArrowLine(24,47)(40,60)
\put(-6,60){$\mu$} 
\put(45,60){$e$} 
\ArrowLine(0,-10)(20,0) 
\ArrowLine(20,0)(40,-10)
\ArrowArc(20,32)(15,100,430)
\put(37,31){$f_1$} 
\put(-16,-12){$f_2$} 
\put(46,-10){$f_2$} 
\Photon(20,17)(20,0){2}{3} 
\end{picture}
\hspace{2cm}
\begin{picture}(70,70)
\ArrowLine(0,60)(20,30)
\ArrowLine(20,30)(40,60)
\put(-6,60){$\mu$} 
\put(45,60){$e$} 
\GCirc(20,25){5}{.7}
\ArrowLine(0,-10)(20,20) 
\ArrowLine(20,20)(40,-10)
\put(-16,-12){$f_2$} 
\put(46,-10){$f_2$} 
\Photon(10,45)(30,45){2}{3} 
\end{picture}
\hspace{2cm}
\begin{picture}(70,70)
\ArrowLine(0,60)(20,30)
\ArrowLine(20,30)(40,60)
\put(-6,60){$\mu$} 
\put(45,60){$e$} 
\GCirc(20,25){5}{.7}
\ArrowLine(0,-10)(20,20) 
\ArrowLine(20,20)(40,-10)
\put(-16,-12){$f_2$} 
\put(46,-10){$f_2$} 
\Photon(10,5)(30,5){2}{3} 
\end{picture}
\vspace{1cm}

\begin{picture}(70,70)
\ArrowLine(0,60)(20,30)
\ArrowLine(20,30)(40,60)
\put(-6,60){$\mu$} 
\put(45,60){$e$} 
\GCirc(20,25){5}{.7}
\ArrowLine(0,-10)(20,20) 
\ArrowLine(20,20)(40,-10)
\put(-16,-12){$f_2$} 
\put(46,-10){$f_2$} 
\Photon(10,45)(10,5){2}{4} 
\end{picture}
\hspace{1.5cm}
\begin{picture}(70,70)
\ArrowLine(0,60)(20,30)
\ArrowLine(20,30)(40,60)
\put(-6,60){$\mu$} 
\put(45,60){$e$} 
\GCirc(20,25){5}{.7}
\ArrowLine(0,-10)(20,20) 
\ArrowLine(20,20)(40,-10)
\put(-16,-12){$f_2$} 
\put(46,-10){$f_2$} 
\Photon(30,45)(30,5){2}{4} 
\end{picture}
\hspace{1.5cm}
\begin{picture}(70,70)
\ArrowLine(0,60)(20,30)
\ArrowLine(20,30)(40,60)
\put(-6,60){$\mu$} 
\put(45,60){$e$} 
\GCirc(20,25){5}{.7}
\ArrowLine(0,-10)(20,20) 
\ArrowLine(20,20)(40,-10)
\put(-16,-12){$f_2$} 
\put(46,-10){$f_2$} 
\PhotonArc(20,25)(15,115,290){2}{5} 
\end{picture}
\hspace{1.5cm}
\begin{picture}(70,70)
\ArrowLine(0,60)(20,30)
\ArrowLine(20,30)(40,60)
\put(-6,60){$\mu$} 
\put(45,60){$e$} 
\GCirc(20,25){5}{.7}
\ArrowLine(0,-10)(20,20) 
\ArrowLine(20,20)(40,-10)
\put(-16,-12){$f_2$} 
\put(46,-10){$f_2$} 
\PhotonArc(20,25)(15,290,115){2}{5} 
\end{picture}
\end{center}
\caption{ Examples of one-loop gauge vertex corrections to 4-fermion operators.
The first two  diagrams are the penguins.
The last six diagrams  contribute to operator
mixing and running, but can  only change the Lorentz or gauge structure
of the operators, not the flavour structure.
Missing are the wave-function renormalisation diagrams; for $V \pm A$
Lorentz structure in the grey blob, this cancels diagrams 3 and 4.
 \label{fig:diag1}
}
\end{figure}
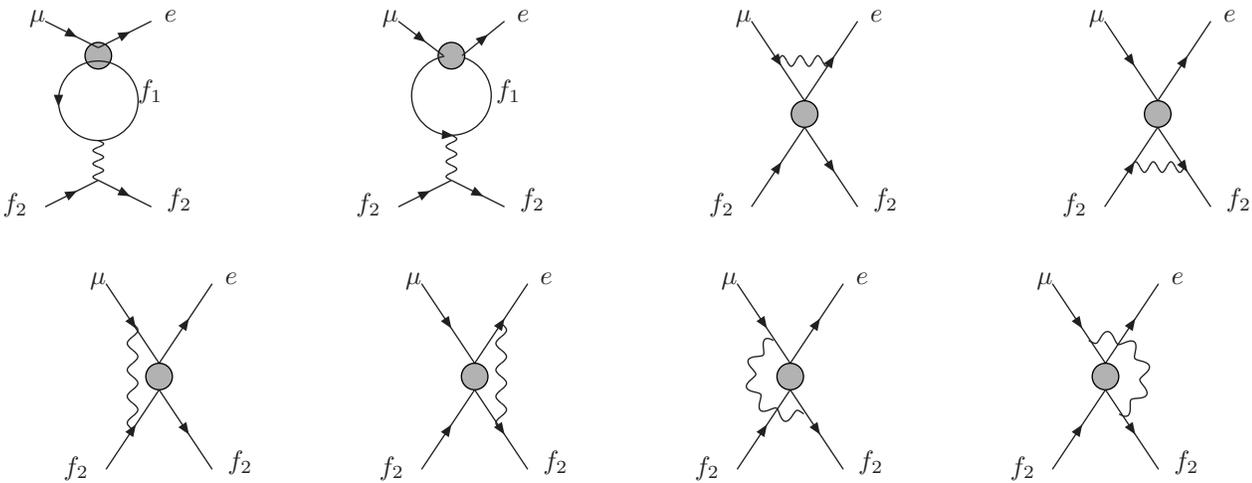

 QCD running concerns the two-lepton-two-quark operators,
and the two-lepton-two-gluon operators. 
The gluon operators are neglected here, because  
they do not contribute at LO to $\meg$,
which is the example considered in section
\ref{sec:meg}, and because one-loop
matching (not performed here) seems
indicated in order to correctly
account for these operators.
The vector two-quark-two-lepton  operators  do not renormalise
under QCD, because
the quark vector currents are conserved:
that is, diagram 4 of figure \ref{fig:diag1},
with $f_2$ a quark and  the photon replaced by a 
gluon, cancels against the wave-function renormalisation.
  However the same diagram, plus wave-function
renormalisation, causes  the scalar
operators run like masses in QCD ($\gamma^s = 6C_F$):
\beq
  C_{S,XY}^{e \mu qq} (m_W) =
  C_{S,XY}^{e \mu qq} (m_q) 
\left[\frac{ \alpha_s(m_W)}{\alpha_s(m_q)}\right]
^{\frac{4}{\beta_0}} =    C_{S,XY}^{e \mu qq} (m_q) \frac{m_q(m_W)}{m_q(m_q)} 
\label{runmass}
\eeq
for $q \in \{u,d,s,c,b\}$ and $X,Y \in \{L,R\}$, 
so  I follow
\cite{CKOT} in normalising the coefficients with  running quark masses
as after the last equality. However,
for the light quarks ($u,d,s$),
the QCD
running  is  stopped at $\mu \simeq m_\tau$, 
that is, $\alpha_s(m_q)$ is replaced by
$\alpha_s(m_\tau)$.
 Finally, diagram 4
vanishes for  the
tensor four-fermion operators, but
the wave-function diagrams cause the
tensor operators to run as:
\beq
  C_{T,XX}^{e \mu qq} (m_W) =
  C_{T,XX}^{e \mu qq} (m_\tau) 
\left[\frac{ \alpha_s(m_W)}{\alpha_s(m_\tau)}\right] 
^{\frac{-4/3}{\beta_0}} 
~~.
\eeq

In QED running, the vector operators  mix among themselves, 
but they have no mixing into or from the scalars and tensors.
The scalars renormalise themselves and mix to tensors and in some cases to  the
dipoles, and
the tensors renormalise themselves and mix to scalars and  dipoles
(which are dimension 5, so do not mix to other operators).
So the anomalous dimension matrix can be  written: 
\beq
\bm{\Gamma}= \left[
\begin{array}{cc}
\bm{\Gamma}_V & 0\\
0 & \bm{\Gamma}_{STD}
\end{array}
\right]
\label{Gamma}
\eeq
with
\beq
\bm{\Gamma}_{STD}= \left[
\begin{array}{cccccccccccc}
\g_{S,S}^{q_i,q_j} &0 &0  &0 &0 &0 &\g_{S,T}^{q_i,q_j} &0 &0 &0 &0 &0\\
0&\g_{S,S}^{u,u} &0 &0 &0 &0  &0 &\g_{S,T}^{u,u} &0 &0 &0&0 \\
0 &0 &\g_{S,S}^{d,d} &0 &0  &0  &0&0 &\g_{S,T}^{d,d} &0 &0&0 \\
0&0 &0 &\g_{S,S}^{\tau,\tau} &0 &0  &0  &0&0 &\g_{S,T}^{\tau,\tau} &0&0 \\
0&0 &0 &0 &\g_{S,S}^{\mu,\mu} &0  &0 &0 &0 &0 &\g_{S,D}^{\mu,} &0\\
0 &0  &0  &0  &0 &\g_{S,S}^{e,e} &0 &0 &0 &0 & \g_{S,D}^{e,}&0\\
\g_{T,S}^{q_i,q_i} &0  &0 &0 &0  &0 &\g_{T,T}^{q_i,q_i} &0 &0  &0 &\g_{T,D}^{q_i,~} &0\\
0 &\g_{T,S}^{u,u}  &0 &0&0 &0  &0 &\g_{T,T}^{u,u}  &0 &0 &\g_{T,D}^{u,~} &0\\
0 &0  &\g_{T,S}^{d,d} &0&0 &0  &0 &0 &\g_{T,T}^{d,d}   &0 &\g_{T,D}^{d,~} &0\\
0 &0  &0 &\g_{T,S}^{\tau,\tau} &0 &0  &0 &0 &0 &\g_{T,T}^{\tau,\tau}   &\g_{T,D}^{\tau,~} &0\\
0&0 &0 &0 &0 &0  &0 &0 &0 &0 &{\color{black} \g_{D,D}} &0\\
0&0 &0 &0 &0 &0  &0 &0 &0 &0 &0 &0\\
\end{array}
\right]
\label{GammaSTD}
\eeq
where the first super- and sub-script  on the $\g$ submatrices
belongs to the coefficient labelling the row, and the second
indices identify the colomn. Section \ref{sec:meg}
 runs  up the dipole coefficient, for which the 
matrix   $\bm{\Gamma}_V$ is not required, so 
it will given in a subsequent publication.

For QED mixing of four-fermion operators among themselves and to the  dipole,
the relevant diagrams are in figure \ref{fig:diag1}, where the
gauge boson is the photon,
and $f_2 \in \{ u,d,s,c,b,e,\mu,\tau\}$.  These diagrams
allow to compute the $\g$-submatrices of eqn (\ref{Gamma}).
The results are  given in Appendix \ref{sec:anomdim}.
 For the second diagram of figure \ref{fig:diag1},   $f_1 =e,\mu$, 
because  Fiertz transformations were used to obtain
a basis where the $\mu-e$ flavour change is within a spinor contraction.

\section{ At $m_W$}
\label{@mW}

Above the ``intermediate'', weak  scale  of the EFT, 
$m_W \simeq m_h \simeq m_t$,
two things differ with respect to the
low energy theory: the theory and non-renormalisable  operators should 
now respect the weak SU(2)  symmetry, and the particle content
is extended to include the weak gauge bosons,
the higgs, and the top. The additional
requirement of SU(2) invariance will reduce the number
of possible four-fermion operators, whereas
 adding new degrees of freedom ($h,W,Z,t$)
allows more   flavour-changing operators involving
only two fermions.

\subsubsection{Neglecting dimension eight operators}

The EFT above $m_W$ is an expansion in the inverse New Physics
scale $1/M$, where the lowest order  operators
that are  lepton flavour-changing, but  number-conserving,
appear at dimension six; they are listed in appendix \ref{BWP}.
It is  convenient to neglect the next order operators,
which would appear at dimension eight,
because they are numerous, and their RGEs are unknown. 
So it is interesting to explore how small must be the ratio
$v/M$, to justify a parametrisation using dimension six operators.

This question was studied for $\meg$ 
in a 2 Higgs Doublet Model(2HDM) with LFV \cite{meghdm}.
Naively, the dimension eight operators are suppressed
by $v^2/M^2\equiv z$. However, two  enhancements
arise:\\
1) in some cases, the dimension six  and eight
contributions arise at the same loop order, but the
dimension six part is from matching, whereas the 
dimension eight term arises in running and 
is log$^2$-enhanced.  The ratio of dimension six to
eight is then $\propto z\ln^2 z$, which is $\sim .2$ for $M \sim 10 v$.\\
2) The couplings of the New Physics are unknown,
and could have steep hierarchies.  In the 2HDM, 
the  heavy Higgs couplings to light fermions
can be  ${\cal O}(1)$, rather than of order
the fermions' SM  Yukawa coupling.  This
increase is parametrised  in the 2HDM
by $\tan \beta$, which   in some cases 
enhances the dimension eight operators
 with respect to dimension six. In some 2HDMs,
$\tan \beta \lsim 50$,  which I take as a reasonable
estimate of the possible hierarchy of couplings
between dimension six and either operators.

In a generic model,  these two enhancements could combine,
and other sources of enhancement could perhaps arise. 
So  I impose that $M \gsim 20$ TeV ($\sim 100v$), in the hope that
this suppresses dimension eight operators in many models.

\subsection{Tree matching onto  SU(2) invariant operators}
\label{ssec:matchmW}

The  coefficients of 
the  four-fermion operators from below $m_W$, given in  eqns (\ref{obsops}) and
(\ref{opstauW}),  should be matched at $m_W$
 onto  the  coefficients of the SU(2)-invariant  BWP basis,
which are listed  in  appendix \ref{BWP}.  The coefficients on the left
of the equalities   are from below $m_W$, the coefficients on the right are
SU(2)-invariant.  Both sets of coefficients should
be evaluated at $m_W$, and the fermion masses
which appear in the matching conditions
should also be evaluated at $m_W$.

\subsubsection{dipoles}
Above $m_W$,
there is a  dimension six   dipole  operator
for hypercharge,  and another  one for SU(2). They are  given in eqns
 (\ref{magmo2L},\ref{magmo1L}). The  coefficient
of the photon dipole operators 
${\cal O}^{e \mu }_{D,R}$ ,{\color{black} ${\cal O}^{e \mu }_{D,L}$}
from below $m_W$
are the   linear combinations (the photon is 
$A^\a = c_W B^\a + s_W W^\a_3$, and the negative sign is from $\tau_3$):
$$ C^{e\mu}_{D,R} = c_W C^{e\mu}_{eB} -s_W  C^{e\mu}_{eW} ~~~,~~~
  C^{e\mu}_{D,L}  = c_W C^{ *\mu e}_{eB} -s_W  C^{ *\mu e}_{eW}~~~.$$ 
However,  rather than using 
the Hypercharge and SU(2) dipoles 
${\cal O}^{e \mu }_{eB}$ and ${\cal O}^{e \mu }_{eW}$, 
I follow \cite{PS}, and use the
photon and $Z$ dipoles above $m_W$,
merely changing the names of
the photon dipole coefficient 
\bea  
 C^{ij}_{D,R}=  C^{ij}_{e \g} = c_W C^{ij}_{eB} -s_W  C^{ij}_{eW}   
&,& C^{ij}_{e Z}  = -s_W C^{ ij }_{eB} -c_W  C^{ ij }_{eW} 
\label{Dmatch}
 \eea
where $ij \in \{ e\mu, \mu e \}$.
(Notice that since  the operators
are $ {\cal O}^{e\mu}_{D,R }$ and ${\cal O}^{\mu e}_{D,R }$,
the $+.h.c.$ gives the  $ {\cal O}^{ij}_{D,L }$.)

\subsubsection{four-lepton operators}

The 
BWP basis contains only  the ``vector'' 
 four-lepton operators 
 given in eqns (\ref{OLL},\ref{EL},\ref{ELemu}).  
There are also new dimension six interactions of
the $W,Z$ and $h$, described by the operators  of eqns
(\ref{LLpenguin},\ref{LL3penguin},\ref{EEpenguin},\ref{yuk6L}),
which will contribute to four-lepton operators below $m_W$
in matching out the $Z$ and $h$.

There are  a few curiosities
related to the flavour index structure
below and above $m_W$. First,  since the basis below $m_W$
was defined with the $e$-$\mu$ indices
inside a spinor contraction, there
 is a scalar operator from below
 $m_W$ which must be Fiertzed as
given in eqn (\ref{scalarL}). 
Also, there are more distinct flavour structures
for   operators constructed 
with SU(2) doublets,  than singlets:
the SU(2)-invariant operators ${\cal O}^{e \mu ff}_{L L} $ and 
${\cal O}^{e ff \mu}_{L L} $,  both match onto
the below-$m_W$ operator ${\cal O}^{e \mu ff}_{L L} $. However
the  two SU(2) operators are distinct\footnote{ the first
contracts a flavour-changing neutral current to
a flavour-conserving neutral current. The second
contracts two flavour-changing neutral currents,
or can be fiertzed to make one current flavour-conserving
but then both currents are charge-changing (see eqn \ref{OLL})}
for $f =\tau$, but not for $f = e,\mu$.

The coefficients of operators from below $m_W$
(on the left of the equalities) can be
matched  as follows  onto  the SU(2)-invariant
coefficients  to the right:
\bea 
C^{e \mu ll}_{RR}& =& C^{e \mu ll}_{EE} - 2 C^{e \mu}_{HE} g_R^e
\label{4fdebut}\\
C^{e \mu \tau \tau}_{RR}& =& C^{e \mu \tau \tau}_{EE} -  C^{e \mu}_{HE}  g_R^e
\label{19}\\%
C^{e \mu \ell \ell}_{LR} &=& C^{e \mu \ell \ell}_{LE} 
 - (  C^{e \mu}_{HL,3}
+ C^{e \mu}_{HL,1})g_R^e
\label{20}  \\
C^{e \mu \ell \ell}_{RL}& =& C^{\ell \ell e \mu }_{LE} -  C^{e \mu}_{HE} g_L^e \label{21}\\
C^{e \mu \tau \tau}_{LL} & =&  C^{e \mu \tau \tau}_{LL} + C^{e \tau \tau \mu}_{LL}
-(  C^{e \mu}_{HL,3}
+ C^{e \mu}_{HL,1}) g_L^e
\label{22} \\
C^{e \mu ll}_{LL}& =& C^{e \mu ll}_{LL}  -2(  C^{e \mu}_{HL,3}
+ C^{e \mu}_{HL,1}) g_L^e \label{23}\\
&&\nonumber\\
C^{e \mu \ell \ell}_{S,RR}& =&  - \frac{m_\ell C^{e \mu  }_{EH} v}{m_h^2}\label{24} \\
C^{e \mu \tau \tau}_{S,LR}& =& -2 C_{LE}^{\tau \mu e \tau} - \frac{m_\tau C^{ \mu e *}_{EH} v}{m_h^2}\\
C^{e \mu \tau \tau}_{S,RL}& =& -2 C_{LE}^{e \tau \tau \mu}
 - \frac{m_\tau C^{e \mu }_{EH} v}{m_h^2} \\
C^{e \mu \ell \ell}_{S,LL}& =&  - \frac{m_\ell C^{ \mu e * }_{EH} v}{m_h^2}
\label{27}\\
&& \nonumber\\
C^{e \mu \tau \tau}_{T,RR} &= &0 \label{28} \\
C^{e \mu \tau \tau}_{T,LL} &=& 0 \label{29}
\eea
where $\ell \in \{e,\mu,\tau \}$, $l \in \{e,\mu \}$, $s_W = \sin \theta_W$,
and the
Feynman rule for $Z$ couplings to leptons is
$i \frac{g}{2 c_W}(g^e_L P_L + g^e_R P_R)$, with
\beq
 g_R^e = -2 s_W^2 ~~~,~~~ g_L^e = 1-2 s_W^2~~.
\label{eqnubi}
\eeq
  In case of vector operators involving three muons or three
electrons of the same chirality, there can be
two $Z$-exchange diagrams ($u$ and $t$ channel),
which  can give a 2 with respect  to operators involving
$(\bar{e}\mu) (\bar{\tau}\tau)$.
From eqn (\ref{28},\ref{29}),
 the tensor coefficients  vanish at tree-matching. Nonetheless, these
operators are important below $m_W$, 
because as seen in the previous section, scalars
mix to tensors, and tensors to the dipole. 

\subsubsection{two-lepton-two-quark operators}

 Two issues  about the CKM matrix $V$ arise in
matching operators involving quarks at $m_W$: 
does $V$ appear in the coefficients above
or below $m_W$, and are the quark doublets
in the $u$ or $d$ mass basis?
I put $V$ in the coefficients above $m_W$,
because  the experimental
constraints are being matched ``bottom-up''
onto operator coefficients. So  one coefficient
from below $m_W$  will match onto a
sum  of coefficients above $m_W$, weighted
by  CKM matrix elements.  Secondly,  the quark doublets above $m_W$
are taken in the $u,c,t$ mass eigenstate basis,
because it is convenient for
translating up in scale the bound on $\meg$,
as will be done  in section \ref{sec:meg}.
This is because  tensor operators 
 mix to the dipoles, and  only for $u$-type
quarks are there  SU(2) invariant
dimension six  tensors operators.

The BWP basis of { two-lepton-two-quark operators} has seven vector
operators
given in eqns (\ref{OLQ1},\ref{OLQ3},\ref{OEQ},\ref{OLU},\ref{OLD},\ref{OEU},\ref{OED})
and two scalars and a tensor given in eqns(\ref{scalaremu},\ref{scalarmue},\ref{scalarDemu},\ref{scalarDmue},\ref{tensoremu},\ref{tensormue}). 
For the first two generations and the $b$ quark,
 the coefficents from below $m_W$ (left side of equality)
can be matched to the coefficients  above $m_W$ as follows:
\bea
C^{e \mu u_n u_n}_{LL}& =&
C^{e \mu nn}_{LQ (1)} -  C^{e \mu nn}_{LQ (3)} - g^u_L
( C^{e \mu }_{HL  (1)} + C^{e \mu }_{HL  (3)} )   \\
C^{e \mu d_n d_n}_{LL}& =& 
\sum_{jk} V_{jn} V^*_{kn} (
C^{e \mu jk}_{LQ (1)} +  C^{e \mu jk}_{LQ (3)}) 
 - g^d_L ( C^{e \mu }_{HL  (1)} + C^{e \mu }_{HL  (3)} ) 
\\
C^{e \mu u_n u_n}_{RR}& =& C^{e \mu nn}_{EU} - g^u_R  C^{e \mu }_{HE}\\
C^{e \mu d_n d_n}_{RR}& =& C^{e \mu nn}_{ED} - g^d_R  C^{e \mu }_{HE}\\
C^{e \mu u_n u_n}_{LR}& =& C^{e \mu nn}_{LU}  - g^u_R ( C^{e \mu }_{HL  (1)} + C^{e \mu }_{HL  (3)} ) \\
C^{e \mu d_n d_n}_{LR}& =&  C^{e  \mu  nn}_{LD}- g^d_R ( C^{e \mu }_{HL  (1)} + C^{e \mu }_{HL  (3)} )\\
C^{e \mu u_n u_n}_{RL} 
& =& C^{e \mu nn}_{EQ}  - g^u_L  C^{e \mu }_{HE}\\
 C^{e \mu d_n d_n}_{RL}& =& 
\sum_{jk} V_{jn} V^*_{kn} 
C^{e \mu jk}_{EQ}  - g^d_L  C^{e \mu }_{HE}\\
&&\nonumber \\
C^{e \mu u_nu_n}_{S,LL}& =& C_{LEQU}^{* \mu e nn} -  \frac{m_{u_n} v}{m_h^2}  C^{ \mu  e *}_{EH} \\
C^{e \mu d_n d_n}_{S,LL}& =& -  \frac{m_{d_n} v}{m_h^2}  C^{ \mu  e *}_{EH} \label{40}\\
C^{e \mu u_n u_n}_{S,RR}& =& C_{LEQU}^{e \mu  nn} -  \frac{m_{u_n} v}{m_h^2}  C^{ e \mu   }_{EH}\\
C^{e \mu d_n d_n}_{S,RR}& =& -  \frac{m_{d_n} v}{m_h^2}  C^{e  \mu   }_{EH}\label{42}\\
C^{e \mu u_n u_n}_{S,LR}& =&   -  \frac{m_{u_n} v}{m_h^2}  C^{  \mu e *  }_{EH}\\
C^{e \mu d_n d_n}_{S,LR}& =& C_{LEDQ}^{* \mu e nn}  -  \frac{m_{d_n} v}{m_h^2}  C^{  \mu e *  }_{EH}\\
C^{e \mu u_n u_n}_{S,RL}& =&  -  \frac{m_{u_n} v}{m_h^2}  C^{ e \mu   }_{EH}\\
C^{e \mu d_n d_n}_{S,RL}& =& C_{LEDQ}^{e \mu  nn} -  \frac{m_{d_n} v}{m_h^2}  C^{ e \mu   }_{EH}\\
C^{e \mu u_n u_n}_{T,LL}& =& C_{T,LEQU}^{*  \mu e nn}\\
C^{e \mu d_n d_n}_{T,LL}& =& 0\label{48}\\
C^{e \mu u_n u_n}_{T,RR}& =& C_{T,LEQU}^{e  \mu  n n}\\
C^{e \mu d_n d_n}_{T,RR}& =& 0 \label{50}
\label{Tfin}
\eea
where $u_n \in \{u,c \}$, $d_n \in \{d,s,b \}$,
and
\beq
g^u_L = 1-\frac{4}{3} s_W^2~~,~~
g^u_R = -\frac{4}{3} s_W^2~~,~~
g^d_L = -1+\frac{2}{3} s_W^2~~,~~
g^d_R = \frac{2}{3} s_W^2~~~.
\eeq

\subsection{Comments on tree matching}
\label{ssec:comments}

One observes that the consequences  of matching at $m_W$,
at tree level, 
are different for vector vs scalar-tensor-dipole operators.
In the vector case, there are more operator
coefficients in the SU(2)-invariant theory above $m_W$
than in the QCD$\times$QED-invariant theory below, 
whereas there are fewer for the scalar-tensor operators.
This means that  
SU(2)-invariance 
should predict  some correlations in
 the  scalar-tensor coefficients
 below $m_W$.  Whereas, if one was trying
to reconstruct the coefficients of the
SU(2)-invariant operators from data,
some additional input ({\it e.g.} from $Z$ physics,
 neutrino interactions\cite{Falkowski}, or
loop matching) would be
required for the vector operators, beyond the
coefficients of  the operators of eqns
(\ref{obsops}) and (\ref{opstauW}).

\subsubsection{ The vector operators}

Consider first the vector operators,  including the ``penguin''
operators of Eqns (\ref{LLpenguin},\ref{LL3penguin},\ref{EEpenguin}) as well as the four-fermion operators.
\bit 
\item  In the
case of four-lepton operators with
flavour indices
$e \mu ee$ or $e\mu\mu \mu$, 
there are the same number of independant
coefficients above and below. There
is one extra four-lepton operator above $m_W$
for flavour indices  $e\mu\tau \tau$,
as can be seen from eqn (\ref{22}).
\item 
There are fewer two-lepton-two-quark
operator coefficients above $m_W$ than below.
It is clear that the  operators
${\cal O}_{LR}^{e \mu qq}$, ${\cal O}_{RR}^{e \mu qq}$
from below $m_W$ with $q \in \{u,d,s,c,b\}$
are equivalent to the 
${\cal O}_{LU}^{e \mu u_nu_n}$,
${\cal O}_{LD}^{e \mu d_nd_n}$,
${\cal O}_{EU}^{e \mu u_nu_n}$,
${\cal O}_{ED}^{e \mu d_nd_n}$
operators from above. And that the 
${\cal O}_{LL}^{e \mu qq}$ from below $m_W$
 with $q \in \{u,d,c,s,b\}$ have the same
number of independent coefficients
as ${\cal O}_{LQ (1)}^{e \mu nn}$  and
 ${\cal O}_{LQ (3)}^{e \mu nn}$. The restriction
occurs  between  
${\cal O}_{RL}^{e \mu qq}$
from below $m_W$,  where there
are five coefficients corresponding to $q \in \{u,d,s,c,b\}$,
and ${\cal O}_{EQ}^{e \mu nn}$ above $m_W$,
which has a coefficient per generation. 
Neglecting CKM sums, this suggests that SU(2) predicts
the  $C_{RL}^{e \mu uu} - C_{RL}^{e \mu dd}=0$
and $C_{RL}^{e \mu cc} - C_{RL}^{e \mu ss} = 0$;
however, there is a  penguin operator
which contribute to both differences, so
only the difference of differences is
an SU(2) prediction (possibly blurred by CKM).

\item The ``penguin'' operators  from above
$m_W$(see eqns (\ref{LLpenguin},\ref{LL3penguin},\ref{EEpenguin})) 
give the $Z$ a vertex with $\bar{e} \g  P_Y \mu$,
which matches onto $(\bar{e} \g P_Y \mu)
(\bar{f} \g P_X  f)$ operators for all
the SM fermions below $m_W$, in ratios
fixed by the SM $Z$ couplings.  This
contribution adds to  the 
four-fermion operator induced at
the scale $M$ in the EFT, as  given
in the matching conditions eqns(\ref{4fdebut}-\ref{Tfin}).  
So  the coefficient of the  $\bar{e} \Zslash  P_R \mu$
penguin operator of eqn (\ref{EEpenguin}) 
could be determined from 
 $C_{RL}^{e \mu uu} - C_{RL}^{e \mu dd}$,
as discussed in the item above.
The coefficients
of the  two remaining  penguin operators
are ``extra'':  in naive coefficient-counting,
there are two more  
vector  coefficients  above $m_W$ than below. 
However, they  are not
completely ``free'', because  they would match at one-loop onto the photon dipole operator at $m_W$.

 These extra penguins  are related to   the common
wisdom,  that it is interesting for  
ATLAS and CMS  to look
for $\Ztm$ and $\Zte$ decays,
but that they are unlikely to see $\Zme$\cite{ATLASLFV}. 
The point\cite{DLV} is that an interaction
  $\bar{\tau} \Zslash  \mu$ would contribute
at tree level to $\tau \to \mu \bar{l}l$, and
at one loop to $\tmg$.
To be
within the sensitivity of the LHC,
the coefficient of this coupling needs
to exceed the naive bound from  $\tau \to \mu \bar{l}l$.
 However,  
 $BR(\tlll)$ \cite{t3l}  is
controlled by    coefficients $C_{XY}^{\mu \tau ll}$, $C_{YY}^{\mu \tau ll}$,
analogous to the coefficients on the
left of eqns (\ref{4fdebut}-\ref{23}),
which are the sum of SU(2)-invariant   
four-fermion and  penguin
coefficients. So the penguin
coefficient could exceed the expected
bound from $\tlll$, 
provided that it is tuned against the
four-fermion coefficient \footnote{Of course, since
the penguin contributes to all
four-fermion operators $(\bar{\mu} \g \tau)
(\bar{f} \g f)$,  the coefficients of many
other operators might need to be tuned
against the penguin too. An apparently less contrived
way to engineer this, is to use the equations of motion
to replace the penguin operator by a derivative
operator $\partial_\a Z^{\a \b} \bar{\mu} \g_\b \tau$\cite{DMEFT},
which is suppressed at low
energy by the $Z$ four-momentum.}. 
This same argument could apply to 
a  $\bar{e} \Zslash  \mu$ coupling
and the bound from $\meee$, although
more tuning would be required, since
the bound on $\meee$ \cite{meee} is more restrictive. 
However, the $Z$ penguins also contribute
at one-loop to  $\meg$ and $\tmg$.
And whereas the experimental constraint
on  $\tmg$ \cite{tlg}  is consistent with $\Ztm$
being detectable at the LHC, the
bound from
$\meg$ implies that 
a $\bar{e} \Zslash   \mu$ interaction,
with coefficient of a magnitude 
that the LHC could detect, would
overcontribute to $\meg$ by several 
orders of magnitude \cite{DLV}. 
\eit

\subsubsection{The scalar, tensor, and dipole operators}

\bit
\item  Above $m_W$, there are two dipoles,
given in eqn (\ref{Dmatch}).
At tree-level, the $Z$-dipole
does not match onto any operator below $m_W$.
\item 
  There are  no
dimension six,  SU(2)-invariant four-fermion operators 
to match onto  the tensor operators ${\cal O}^{e\mu ff}_{T,YY}$
for $f \in \{ \tau,d,s,b\}$.  Furthermore,
in tree level matching, the tensors are not
generated by any heavy particle exchange.
They are presumeably generated in one-loop
matching by the same diagrams that give 
the mixing below $m_W$, but this should
be subdominant because lacking the log.
\item  There are  no
dimension six,  SU(2) invariant four-fermion operators 
to match onto the scalar operators
${\cal O}^{e\mu u_nu_n}_{S,YX}$ and
${\cal O}^{e\mu ff}_{S,YY}$ for 
 $f \in \{ e,\mu,\tau,d,s,b\}$, $u_n \in \{ u,c\}$
and $X\neq Y$. However,  SM Higgs
exchange, combined with the  $H^\dagger H \bar{L}HE$
operator,  will generate  these operators in tree matching,
weighted by   $m_f v/m_h^2$ or
 $m_{u_n} v/m_h^2$.
So it is a tree-level SU(2) prediction that
these coefficients are small, as noted by
\cite{AGC}. Since the  coefficients of scalar operators involving quarks
are normalised by a running quark mass, see eqn (\ref{runmass}),
one obtains $C^{e\mu ff}_{S,...} (m_\tau) \simeq  -C^{e\mu}_{EH} (m_W) m_f(m_\tau) m_t/m_h^2 $.

\eit

\subsubsection{Matching at ``Leading'' Order}
\label{ssec:LO}

The aim of a  bottom-up EFT analysis  
is to translate the bounds from  several observables 
to combinations of operator
coefficients at the high scale.
So one must compute    the numerically  largest contribution
of each operator to several observables 
($\meg$, $\mec$ and $\meee$, in the case of $\mu$-$e$ flavour change).  
It is interesting to
have constraints  from different observables,
rather than just the best bound, because
 there are more operators than observables,  
so a weaker constraint on a different combination
of coefficients can  reduce degeneracies.
However, in this paper, only the experimental
bound from $\meg$  is considered, so
the aim is to obtain the best bound 
it sets on all
operator coefficients. 

In the next section, we will see that  tree matching and
one-loop running, as performed so far, do not reproduce the correct
constraints from $\meg$   on the operators which parametrise 
LFV interactions of the Higgs and $Z$; that is, 
the  numerically dominant  contributions of these operators
to $\meg$ are not included. In addition, two-loop
QED running \cite{PepeGian} is required below $m_W$ to obtain bounds on 
vector operators. So its clear that the
simplistic  formalism  given here, of tree matching and
one-loop running, does not work for $\meg$. 

It would be interesting to construct
a systematic formalism,
 gauge invariant  and renormalisation scheme
independent,  that allows
to obtain the best bound on each operator
from each observable.   I suppose that such 
a formalism corresponds to  ``leading order''.
Notice that leading order is only defined ``top-down'',
because  it describes the contribution of an operator to an observable.
So to construct a LO formalism  for bottom-up EFT, it seems that
one must work top-down, finding the numerically dominant
contribution of each operator to each observable, then
ensuring that the combination of the contributions
from all the operators is scheme independent.

As previously stated, the LO two-loop running
is neglected in this paper. 
However,  some attempt is made
to perform  LO matching at $m_W$, where
the ``LO contribution'' of a coefficient above
the matching scale 
to a coefficient below,
is pragmatically  defined 
as the numerically dominant
term (and not
the lowest order in the loop expansion,
because this may not
be  the numerically dominant contribution
in presence of hierarchical Yukawas).

So, in summary, the ``Leading Order'' matching
performed for $\meg$ in the next
section will consist of the tree equivalences
given in this section, augmented by some
one and two-loop contributions of operators
that do not mix to the dipole. These
loop  contributions are  obtained by
listing all the operators which
do not mix into the dipole above $m_W$,
estimating their matching contribution
at $m_W$, and including it if it
gives an interesting contraint.


\section{Translating the $\meg$ bound to $M > m_W$}
\label{sec:meg}

In this section, the aim is to use the machinery 
developed  in the previous sections to translate the  experimental bound on
$BR(\meg)$ to a constraint on operator coefficients at
the New Physics scale $M$. 

\subsection{Parametrising $\meg$}
\label{ssec:param}

A flavour-changing dipole operator (in the notation of Kuno and Okada\cite{KO})
\beq
{\cal L}_{\meg} = -\frac{4G_F}{\sqrt{2}} m_\mu \left(
A_R \overline{\mu_R} \sigma^{\a\b} e_L F_{\a\b} +
A_L \overline{\mu_L} \sigma^{\a\b} e_R F_{\a\b}\right)
\label{Lmeg}
\eeq
can be added to the SM Lagrangian at a low-scale $\sim m_\mu$,
and gives a branching ratio 
\beq
BR(\meg)= 384 \pi^2 (|A_R|^2 + |A_L|^2) <5.7\times 10^{-13}
\label{BRmeg}
\eeq
where the constraint is from  the MEG experiment\cite{MEG13}.
If $|A_R| = |A_L|$, then $|A_X| <8.6\times 10^{-9}$,
whereas conservatively only allowing for one
coefficient gives the bound   $|A_X| <1.2\times 10^{-8}.$
Translated to the   coefficients of the dipole
operators of eqn (\ref{obsops}), which are
defined  including a muon Yukawa, this  conservative limit 
gives
\beq
C^{e\mu}_{D,X} = \frac{ A_X M^2}{m_t^2}  <1.2\times 10^{-8}
\frac{ M^2}{m_t^2}
\label{expt}
\eeq

It is interesting to estimate  the scale $M$
to which  experiments currently probe.
One can consider  three
possible guesses for the form of the coefficient of the
operator $\bar{\mu} \sigma \cdot F P_X e$:
\beq
c\frac{m_\mu}{M^2}~~~,~~~
c\frac{v}{M^2}~~~,~~~
c\frac{e v}{16 \pi^2 M^2}
\eeq
where $c\lsim 1$ is a  dimensionless
 combination of numerical factors and couplings constants.
The first guess is the Kuno-Okada normalisation of   (\ref{Lmeg}), 
corresponding to the Higgs leg attached to the
muon line, but a tree diagram, and suggests
that the current data probes scales  up to $ \sim 10^{6}$ GeV.
The second guess gives the
 maximum possible scale of $\sim 10^{8}$ GeV ---
however, it supposes the dipole operator
is generated at tree level, with all couplings $\sim 1$.
The final guess takes into account that
the dipole operator is generated at
one-loop with a photon leg, and gives a maximum scale of
$M \lsim  3 \times 10^{6}$ GeV.  
Notice that this guess is very
similar to the Kuno-Okada normalisation
used  to define the dipoles
in this paper:  $e/(16\pi^2) \sim 3 y_\mu$.
The maximum scale is relevant, because it determines
how large can be the logarithm from the
RGEs above $m_W$. I take the third guess with 
\beq
M \lsim 3 \times 10^{6}~{\rm GeV} ~~~ \Rightarrow
~~~ \ln \frac{M}{m_W} \lsim 10~~.
\label{guessscale}
\eeq

It is also interesting to estimate the loop
order  probed by  the current MEG bound. Counting 
$1/(16 \pi^2)$ for a loop (as if couplings$\times$logarithm $\simeq$ 1),
and assuming that  $M\gsim 10$ TeV (beyond
the reach of the LHC), then eqn (\ref{guessscale})
suggests that three-loop effects could be probed.
In section \ref{ssec:5.4},  estimated  bounds are
given on all  the operators
which  MEG can constrain. Four-fermion
operators are defined to be  ``constrainable''  
if their  coefficients $C$
can be bounded   $C \lsim 1$ at a scale
$M \sim 100 m_t$. It turns that all these
operators are within two SM loops
of the dipole.

\subsection{Running up to $m_W$}
\label{ssec:5.2}

Between  $m_W$ and $m_\tau$, 
various operators mix into the dipole, so at
$m_W$, the exptal bound (\ref{expt})
 applies to the  linear combination of
the coefficients given on the
left-hand-side of eqn (\ref{oprun1l}),
when the dummy index B is taken to be a dipole:
\bea
\vec{C}_D(m_\tau)
& =& \vec{C}_D(m_W)   - \frac{\alpha  }{4\pi} \left\{  \sum_{l=e,\mu} \vec{C}_{Sl}(m_W)  \left[ \gamma^l_{SD}\right] \right. \nonumber\\
&& \left. ~~~~~+\sum_{x=q_i,u,d,\tau} \left( \vec{C}^x_T(m_W) -
\frac{\alpha}{8\pi} \log \frac{m_W}{m_\tau} \vec{C}^x_S (m_W)\left[ \gamma^x_{ST}\right] \right) \left[ \gamma^x_{TD}\right] \right\}
\log \frac{m_W}{m_\tau} 
\eea
where $q_i \in \{s,c,b\}.$
The contribution of $ C^{e\mu ee}_{S,LL}(m_W)$
will be neglected, because it is
constrained by $\meee$. 
A linear combination of 
  $ C^{e\mu dd}_{S,LL}(m_W)$
  $ C^{e\mu ss}_{S,LL}(m_W)$, 
and $ C^{e\mu uu}_{S,LL}(m_W)$
contributes to $\mec$, so possibly an
independent constraint from $\meg$  on a different
combination could be interesting. 
However, I neglect these coefficients too,
to avoid strong interaction issues
and   because  in tree matching {\color{black} at $m_W$}, 
 the first two   are
 Yukawa suppressed to irrelevance 
\footnote{Eqns (\ref{27},\ref{40}) show that these
operators arise at $m_W$ by matching out ${\cal O}_{EH}$,
which gives a larger contribution to the
dipole via top and $W$ loops, as given
in eqn (\ref{top})}.
In the following, I focus on the ``left-handed'' dipole
$C_{D,L}^{e\mu}$. The evolution of $C_{D,R}^{e\mu}$
is similar, so for the ``right-handed
dipole'',  only final results and a few non-trivial
differences  are given 
(which arise due to Higgs loops
above $m_W$, where  $Y_e \ll Y_\mu$ is neglected).
One  obtains 
\bea
C^{e\mu}_{D,L}(m_\tau)&\simeq& 
C^{e\mu}_{D,L}(m_W) +
\frac{e}{4\pi^2}\left(  C^{e\mu \mu \mu}_{S,LL}(m_W)
{\color{black}  - 
\frac{8 Q_u N_c m_u }{m_\mu} C^{e\mu uu}_{T,LL}(m_W)
+ \frac{8  m_d }{ m_\mu} C^{e\mu dd}_{T,LL}(m_W)
+ \frac{8  m_s }{ m_\mu} C^{e\mu ss}_{T,LL}(m_W)}
 \right.
  \nonumber\\&&  \left.
 - 
\frac{8 Q_u N_c m_c }{m_\mu} C^{e\mu cc}_{T,LL}(m_W)
+ \frac{8  m_b }{ m_\mu} C^{e\mu bb}_{T,LL}(m_W)
+ \frac{8  m_\tau }{m_\mu} C^{e\mu \tau \tau}_{T,LL}(m_W)
 \right)
  \nonumber\\&& 
+\frac{ e \alpha }{\pi^3} \left(
+ \frac{2  m_\tau }{m_\mu} C^{e\mu \tau \tau}_{S,LL}(m_W)
+ \sum_{q = s,c,b}\frac{2 N_c Q_q^2  m_q }{m_\mu} C^{e\mu qq}_{S,LL}(m_W) \right)
\eea
where the first parenthese is first order
in $\bm{\Gamma}$,
the second parenthese is the second order
scalar$\to$tensor$\to$dipole
mixing,  $Q_q$ is the electric charge,  and the  $\log \frac{m_W}{m_\tau}$
was taken  $\sim 4$. {\color{black} The light quark ($u,d,s$) tensor
contributions  only include  the mixing between $m_W$ and $m_\tau$;
the (non-perturbative)  mixing  between $m_\tau$ and
$m_\mu$ is  difficult to calculate, so neglected.
Due to this uncertainty, the light quark tensors are
neglected after eqn (\ref{plot}).}
With quark masses evaluated
at $m_W$, this gives
\bea
C^{e\mu}_{D,L}(m_\tau)
 &\simeq& 
C^{e\mu}_{D,L}(m_W) 
{\color{black} -.0016 C^{e\mu uu}_{T,LL}(m_W)
+ .0017 C^{e\mu dd}_{T,LL}(m_W)
+ .035 C^{e\mu ss}_{T,LL}(m_W)}
  \nonumber\\&& ~~~
- 1.0 C^{e\mu cc}_{T,LL}(m_W)  + 1.0 C^{e\mu \tau \tau }_{T,LL}(m_W)  
 + 1.8 C^{e\mu bb}_{T,LL}(m_W)    
  \nonumber\\&& 
+ 10^{-3} \left \{
7.6 C^{e\mu \mu \mu}_{S,LL}(m_W)  + 4.6 C^{e\mu \tau \tau}_{S,LL}(m_W) 
~~+1.4 C^{e\mu bb}_{S,LL}(m_W) +1.5 C^{e\mu cc}_{S,LL}(m_W) 
\right\}~~~~
\label{plot}
\eea
where one notices that the 
scalar$\to$tensor$\to$dipole mixing  of the ``heavy''
fermion ($f \in \{\tau,c,b \}$) operators is  of the same magnitude  as
the scalar$\to$dipole mixing of the $\mu$
operator, because the anomalous dimension
mixing tensors to dipoles is large and enhanced  by $m_f/m_\mu$.
This mixing is the EFT implementation of the  two-loop 
``Barr-Zee'' diagrams (see figure \ref{fig:2loop})
 of  the $\tau$, $c$ and $b$: contracting the
scalar propagator of the Barr-Zee diagram to a point
gives a scalar four-fermion operator, then
the photon exchanged between the muon and heavy fermion
makes a tensor operator, then the heavy fermion
lines are closed to give the dipole.

\begin{figure}[ht]
\unitlength.5mm
\begin{center}
\epsfig{file=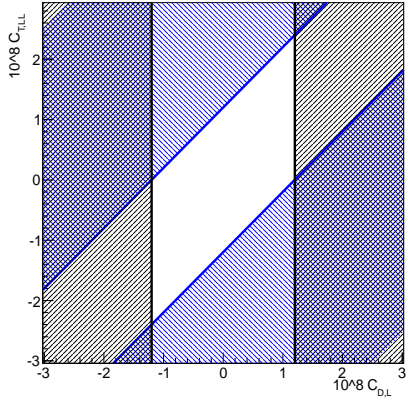,height=6cm,width=6cm}
\hspace{2cm}
\end{center}
\caption{ Between the  vertical black lines is  the
allowed range for the  coefficients of the dipole
operator ${\cal O}_{D,X}^{e \mu}$ (horizontal axis)
and $c$-tensor operator  ${\cal O}_{T,XX}^{e \mu cc}$ (vertical axis), evaluated
at low energy. At $m_W$, the allowed region is between the diagonal 
blue lines, see eqn \ref{plot}. This illustrates that the allowed region 
changes with scale, in this case due to operator
mixing.
\label{fig:c11} }
\end{figure}

At the weak scale, the experimental bound constrains 
a linear combination
of several different operators. It is common to quote the resulting constraints ``one at a time'', that is, retaining only one coefficient in the sum of eqn
(\ref{plot}), and setting the remainder to zero, in order
to obtain a bound.  I will do this later, in  listing
bounds at the scale $M$. However, it is important to remember
that the MEG experiment only  ever gives two constraints  (on
$C^{e\mu}_{D,R}(m_\tau)$ and $C^{e\mu}_{D,L}(m_\tau)$)
in the multi-dimensional space of operator coefficients,
and additions or cancellations are possible among the many 
contributing operators at $m_W$.  This is illustrated
in figure \ref{fig:c11}, where
the black lines give  the experimental bound
at low energy on $C^{e\mu}_{D,L}(m_\tau)$. 
The diagonal black lines are the bound at $m_W$,
in a model  where only the coefficients
$C^{e\mu}_{D,L}(m_W)$ and $C^{e\mu cc}_{T,LL}(m_W)$ are non-zero :
arbitrarily large values of $C^{e\mu}_{D,L}(m_W)$ and $C^{e\mu cc}_{T,LL}(m_W)$
are allowed, provided they are correlated.  
  Including experimental
constraints from $\meee$ and $\muc$ 
would give other constraints
on different linear combinations of coefficients, but the problem 
of having  more operators than
experimental constraints would remain. 

\subsection{Matching at $m_W$}

The tree-level matching conditions of section  \ref{@mW} allow
to translate, at $m_W$,
the coefficients of QCD$\times$QED-invariant operators
  to   SU(2)-invariant coefficients.  With these rules,
 the low-scale dipole coefficient 
can be written
\bea
C^{e\mu}_{D,L}(m_\tau)
&\simeq& 
 C^{\mu e *}_{e \g}(m_W)  -
C^{\mu e cc *}_{LEQU (3)}(m_W)\frac{2 e Q_u   N_c m_c }{m_\mu\pi^2} 
+ C^{\mu e cc *}_{LEQU}(m_W)\frac{2 e  \alpha Q_u^2  N_c m_c }{m_\mu\pi^3} 
\nonumber \\
&&-C^{\mu e *}_{EH}(m_W) \left[ 
\frac{ m_\mu v }{4 \pi ^2m_h^2}
+ \frac{2 e \alpha}{\pi^3}\frac{ m^2_\tau v }{m_\mu m_h^2 }
+ \frac{2  e \alpha   N_c v }{ \pi^3  m_h}
\left(\frac{Q_d^2 m_b^2 + Q_u^2 m_c^2} {m_\mu m_h}\right)
\right] ~~~,
\label{square}
\eea
with a similar equation for $C^{e \mu}_{D,R}(m_\tau)$.
Only four SU(2)-invariant coefficients
are required,  because  for the
leptons and  down-type quarks,
 there are no SU(2)-invariant,
dimension-six  tensor operators,
  nor  scalar operators with the
required $LL$ chiral structure.
The tensor operators are not generated in
matching out the $W,Z,h$ and $t$ at tree level,
so their coefficients can be set
to zero as given in eqns (\ref{28},\ref{29},\ref{48},\ref{50}).
(They could arise in one-loop matching,
via diagrams similar to those giving running below $m_W$,
so the tensor coefficients were retained in the discussion of
the  section \ref{ssec:5.2}.) 
The scalar operators
are generated in matching out the Higgs,
see section \ref{ssec:matchmW}, which gives the square
bracket above.


\begin{figure}[ht]
\unitlength.5mm
\SetScale{1.418}
\begin{boldmath}
\begin{center}
\begin{picture}(60,80)(0,0)
\ArrowLine(0,0)(20,0)
\ArrowLine(20,0)(60,0)
\ArrowLine(60,0)(80,0)
\DashLine(8,45)(28,30){1}
\DashLine(60,0)(48,18){1}
\DashLine(60,0)(63,22){1}
\DashLine(60,0)(78,13){1}
\CArc(40,25)(10,0,360)
\Photon(40,35)(40,60){2}{4}
\Photon(33,18)(20,0){2}{4}
\GCirc(60,0){3}{.7}
\Text(5,-5)[r]{$\mu$}
\Text(82,0)[l]{$e$}
\Text(45,50)[l]{$\gamma$}
\Text(52,35)[l]{$t$}
\Text(19,15)[c]{$\gamma$}
\Text(51,10)[r]{$h$}
\Text(60,-12)[c]{$ C^{\mu e*}_{EH}\frac{ v^2}{M^2}$}
\end{picture}
\qquad\qquad\qquad
\qquad
\begin{picture}(60,80)(0,0)
\ArrowLine(0,0)(20,0)
\ArrowLine(20,0)(60,0)
\ArrowLine(60,0)(80,0)
\DashLine(8,45)(24,32){1}
\DashLine(60,0)(48,17){1}
\DashLine(60,0)(63,22){1}
\DashLine(60,0)(78,13){1}
\PhotonArc(40,25)(10,0,360){2}{7}
\Photon(40,33)(40,60){2}{4}
\Photon(33,17)(20,0){2}{4}
\GCirc(60,0){3}{.7}
\Text(5,-5)[r]{$\mu$}
\Text(82,0)[l]{$e$}
\Text(48,50)[l]{$\gamma$}
\Text(57,35)[r]{$W$}
\Text(16,15)[c]{$\gamma$}
\Text(60,15)[r]{$h$}
\Text(60,-12)[c]{$C^{\mu e*}_{EH}\frac{ v^2}{M^2}$}
\end{picture}
\end{center}
\end{boldmath}
\vspace{10mm}
\caption{The two-loop ``Barr-Zee'' diagrams  which gives
the largest contribution of the  $H^\dagger H \bar{L} HE$
operator to  the dipole  below $m_W$. 
The grey disk is the dimension six
interaction, with  two  Higgs
legs connecting to the vev. The Higgs
line approaching the top loop
indicates a mass insertion  somewhere on
the top loop.
\label{fig:2loop}}
\end{figure}
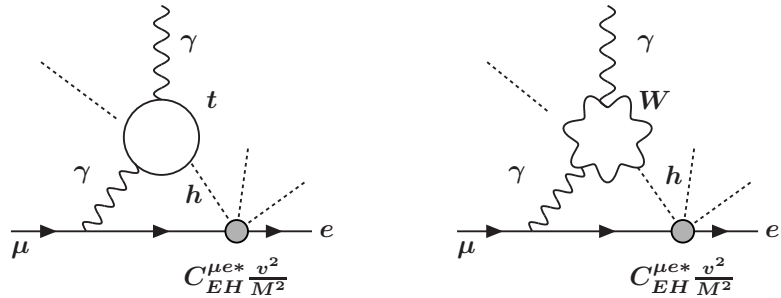

However, it is well-known that this estimate has missed
the largest contribution from  ${\cal O}^{\mu e *}_{EH}$
to the dipole operator below $m_W$, 
which are   ``Barr-Zee'' diagrams  with 
 the SM Higgs and 
a top or $W$ loop, as illustrated in figure \ref{fig:2loop}.
Despite being suppressed by two loops, these diagrams are
enhanced by  $m^2_t/m^2_\mu$ or  $m^2_W/m^2_\mu$.
In SU(2)-invariant notation, these
diagrams generate a ``dimension eight'' dipole
$H^\dagger H \bar{L} H\sigma \cdot F  E$. However,
SU(2) is irrelevant below $m_W$, so this is an ${\cal O}(1/M^2)$
matching contribution to ${\cal O}^{e \mu}_{D,X}$.
 For lack of good ideas
on how to do a well-defined perturbation theory in
many small  parameters (in particular, loops and
hierarchical yukawas), I  retrieve
from the results of Chang,Hou and Keung\cite{CHK},
the evaluation of the Barr-Zee diagrams with a SM Higgs
and a top or W loop (which have opposite sign):
\beq
\Delta C^{e\mu}_{D,L}(m_W)  \simeq -
 C^{\mu e *}_{EH}(m_W)  
\left[ 
\frac{e\alpha }{16\pi^3Y_\mu } \left(  Q_t^2   N_c Y^2_t
-\frac{7}{2}   \right)
\right]  \simeq  C^{\mu e *}_{EH}(m_W)  \left[ 
\frac{e\alpha }{8\pi^3Y_\mu } 
\right]
\label{top}
\eeq
 and substitute the square brackets of eqn (\ref{top})
for those in eqn  (\ref{square}).

Having started  cherry-picking the ``leading'' contributions   
 from higher order,
it is interesting also to include the one-loop matching contribution
of the ``penguin'' operators  of eqns (\ref{LLpenguin},
\ref{LL3penguin},\ref{EEpenguin}). These give a lepton-flavour-changing 
vertex to the $Z$, which contributes  to $\Zme$
and at one-loop to $\meg$.
As discussed in section \ref{ssec:comments}, in the context of
LHC searches for flavour-changing $Z$ decays, $\meee$
give a restrictive bound on
a combination of  the penguins {\it plus} four fermion operators.
So  even if weaker, an  independent constraint from $\meg$, on a different
combination of operators,  is interesting. 
The one-loop diagram with  a flavour-changing
$Z$-penguin vertex, gives   contributions to the dipole coefficients:
\bea
\Delta C^{\mu e *}_{e\g} (m_W)  &\simeq& \frac{e }{16\pi^2}
 g^e_L C^{e \mu}_{HE} (m_W) \nonumber\\
\Delta C^{e \mu  }_{e\g} (m_W)  &\simeq& \frac{e }{16\pi^2}
 g^e_R \left( C^{e \mu}_{HL(1)}  (m_W)
+  C^{e \mu}_{HL(3)} (m_W)
\right) ~~~,
\eea
where $g^e_L, g^e_R$ are given in eqn (\ref{eqnubi}),
no muon Yukawa appears in the matching coefficient
because it is implicit in the dipole operator definition,
and the electron Yukawa was neglected (which is why different
penguins mix into the above two dipoles). 
The contribution $\Delta C^{\mu e *}_{e\g}$ 
is to  be  added to  the right side of eqn  (\ref{square})
, and $\Delta C^{e \mu  }_{e\g}$ 
should be  added to 
the modification of  eqn  (\ref{square}) appropriate
to $C_{D,R}^{e \mu}$.

\subsection{Running up to $M$} 
\label{ssec:5.4}

At $m_W$,  $C^{e \mu}_{D,L} (m_\tau)$ can be written
as a linear combination of
$C^{\mu e *}_{e\g}(m_W)$,  $C^{\mu e *}_{eZ}(m_W)$,   $C^{\mu e cc *}_{LEQU(1)}(m_W)$,
  $C^{\mu e cc *}_{LEQU(3)}(m_W)$, $ C^{e \mu}_{HE}(m_W)$,
and $C^{\mu e *}_{EH}(m_W)$.
 The  RGEs to evolve  these coefficients
up  to $M$
are given in \cite{JMT,PS}, and
 generate more intricate  and extensive operator
mixing than  was present below $m_W$.
The aim here is to present
manageable analytic formulae, that approximate  the 
``leading'' (= numerically most important) 
 constraints on all the   constrainable coefficients
at the scale $M$.
Recall that an operator coefficient was
defined here to be constrainable if
the current MEG bound, as given  in eqn (\ref{expt}), 
implies $C < 1$ at  $M \simeq 100 m_t$.

Consider first  $C^{\mu e *}_{EH}$. Neglecting
its self-renormalisation between $M$ and $m_W$,
because the anomalous dimension $\times
\ln M/m_W < 16 \pi^2$,
 the ``one-operator-at-a-time''
constraint at $M \simeq 100 m_t$ is  $C^{\mu e *}_{EH} \lsim .01$.
 So  there could
be a bound on operators that mix into
${\cal O}^{\mu e *}_{EH}$ in running between
$M$ and $ m_W$. These include 
 the $Z$ and $\g$ dipoles, which
can be neglected here because they  have more 
direct contributions
to $\meg$.   There
is also a $Y_\mu$-suppressed mixing from the
``penguin'' operators, which is neglected
because the penguins match at one loop 
onto the dipole at $m_W$. 
 So I  approximate
\beq
C^{\mu e *}_{EH} (m_W) =   C^{\mu e *}_{EH}(M) ~~.
\eeq

Consider next the penguin operators of eqns
 (\ref{LLpenguin}-\ref{EEpenguin}), which match at
one-loop to the dipole.  The bound on
the coefficient at  
$M \simeq 100 m_t$ is  $C^{e \mu}_{HE} \lsim .1$,
so I neglect mixing into these operators, and 
approximate
\beq
C^{e \mu }_{HE} (m_W) =   C^{e \mu }_{HE}(M) ~~,~~
 C^{e \mu}_{HL(1)} (m_W) = C^{e \mu}_{HL(1)} (M) ~~,~~
C^{e \mu}_{HL(3)} (m_W) =C^{e \mu}_{HL(3)} (M) ~~.
\eeq

In running
from $M \to m_W$, the RGEs given in \cite{PS,JMT}
show that gauge interactions will renormalise
 the photon dipole  coefficient $C_{e\g}^{\mu e *}$, 
and cause it to  receive contributions from   $C_{e Z}^{\mu e *}$,
 $C^{\mu e cc *}_{LEQU(1)}$,   $C^{\mu e cc *}_{LEQU(3)}$,
 $C^{\mu e tt *}_{LEQU(1)}$, and $C^{\mu e tt*}_{LEQU(3)}$.
This  gauge mixing of scalars to tensors to  dipoles
is analogous to the QED mixing below $m_W$.
In addition, as given in \cite{JMT},
Higgs loops will mix  vector four-fermion
operators 
into  scalars and  tensors.  In the following,
the  third order vector$\to$scalar$\to$tensor$\to$dipole mixing is 
neglected, and only the vector$\to$tensor$\to$dipole
is retained for vector and tensor operators
with a top bilinear.

Defining a coefficient vector 
$$
\vec{C} = (C_{EU}^{\mu e tt*}, 
C_{EQ}^{\mu e tt*}, C^{\mu e tt*}_{LEQU(1)},  C^{\mu e cc*}_{LEQU(1)},
C^{\mu e tt *}_{LEQU(3)}, C^{\mu e cc *}_{LEQU(3)},
C_{e\g}^{\mu e *}, C_{eZ}^{\mu e *})
$$
then, from \cite{JMT,PS},
 the electroweak  anomalous dimension matrix $\bm{\gamma}_{\g t}$ such that
$\mu \partial \vec{C}/\partial \mu = \frac{\alpha_{em}}{4\pi} \vec{C} \bm{\gamma}$ is approximately
\bea
\bm{\gamma}_{\g t} \sim \left[
\begin{array}{cccccccc}
0&0&0&0&-\frac{ Y_t Y_\mu}{2e^2}&0&0&0\\
~0~&~0~&0&0&-\frac{ Y_t Y_\mu}{2e^2}&0&0&0\\
0&0&-5+ \frac{15 Y_t^2}{2e^2} &0& \frac{7}{3} &0 
&0 &0\\
0&0&0&-5+ \frac{15 Y_c^2}{2e^2} &0& \frac{7}{3} &0 
&0 \\
0&0& 112 &0& 8.5 + \frac{3Y_t^2}{2e^2}&0&\frac{16Y_t}{eY_\mu} &\frac{8 Y_t}{\sqrt{3}e Y_\mu}\\
0&0&0&112&0 & 8.5 + \frac{3Y_c^2}{2e^2}&\frac{16Y_c}{eY_\mu} &\frac{8 Y_c}{\sqrt{3}eY_\mu}\\
0&0&0 &0&\frac{7Y_tY_\mu}{e}&\frac{7Y_cY_\mu}{e}&
7 +\frac{3Y_t^2}{e^2}
& -\frac{24}{\sqrt{3}}\\
0&0&0&0&\frac{22Y_tY_\mu}{6 e}& \frac{22Y_cY_\mu}{6 e}& \frac{12}{\sqrt{3}} &-\frac{8}{3}+ \frac{3Y_t^2}{e^2}\\
\end{array}
\right]
\label{gamgt}
\eea
where small Yukawa couplings and fractions were neglected,
 $\sin^2\theta_W = 1/4$, and renormalisation and mixing
to the vectors was neglected because they only affect
the dipole at ${\cal O}(\alpha^2 \log^2)$. 
The RGE for the tensor coefficient
$C^{e \mu  tt }_{LEQU(3)}$, which mixes to the
 ``right-handed'' dipole $C_{e\g}^{e \mu }$ would
instead include the vector contribution:
\beq
\mu \frac{\partial }{\partial \mu} C_{LEQU(3)}^{ e \mu tt} = ...-
\frac{\alpha_{em}}{4\pi} \frac{Y_tY_\mu}{2e^2} ( C_{LU}^{\mu e tt} +C_{LQ(1)}^{\mu e tt}- 3C_{LQ(3)}^{\mu e tt}) ~~~,
\label{otherR}
\eeq
rather than the first two rows of eqn(\ref{gamgt}). 
 The approximate solution of these RGEs, 
  if  the running of gauge and Yukawa
couplings is neglected \footnote{including  $\alpha_s$, so the quark   operators
no longer run as a power of $\alpha_s(\mu)$},
is
\bea
   C_{B}(m_W) &\simeq & 
   C_{A}(M)
\left(
\delta_{A,B} - \frac{\alpha_{em}  }{4\pi} \left[ \gamma_{\g t}\right]_{A,B}
\ln \frac{M}{m_W}
 + 
\frac{\alpha_{em} ^2 }{32\pi^2} \left[ \gamma_{\g t} \gamma_{\g t}\right]_{A,B}
\ln ^2  \frac{M}{m_W} +..
\right) ~~~.
\label{77}
\eea
Allowing the index $B$   of eqn  (\ref{77})
to run  over the coefficients
present on the right side of eqn (\ref{square}),    the anomalous
dimension matrix of eqn (\ref{gamgt}) and the bound (\ref{expt}) give
\bea
1.2\times 10^{-8}\frac{M^2}{m_t^2} &\gappeq&
 C^{\mu e *}_{e \g}(M)   -0.016 C^{\mu e *}_{EH}(M)  + 0.001   C^{e \mu}_{HE} (M) 
 -0.0043C^{\mu e *}_{e Z}(M) \ln \frac{M}{m_W}
 \nonumber \\
&&
- 59 C^{\mu e tt*}_{LEQU(3)}(M) \ln \frac{M}{m_W} 
- C^{\mu e cc*}_{LEQU(3)}(M)   \left( 0.43 \ln \frac{M}{m_W} + 1.5\right)
 \nonumber \\
&&+  0.039 C^{\mu e tt*}_{LEQU(1)}(M) \ln ^2  \frac{M}{m_W}   
+ 0.002 \left( 1 +  \ln \frac{M}{m_W}  \right)  C^{\mu e cc*}_{LEQU(1)}(M)
 \nonumber \\
&& - 4.8 \times 10^{-5} \ln^2 \frac{M}{m_W}
{\Big (}   C_{EQ}^{\mu e tt *}(M) +C_{EU}^{\mu e tt *}(M) {\Big )}
\label{bornes}
\eea
(where $m_t$ is written instead of the Higgs vev, to avoid
$\sqrt{2}$ issues).   This constraint, as well as the equivalent bound on 
$C_{e\g}^{e \mu}(m_\tau)$:
\bea
1.2\times 10^{-8}\frac{M^2}{m_t^2} &\gappeq&
 C^{e \mu }_{e \g}(M)   -0.016 C^{e \mu }_{EH}(M)  + 0.001  {\Big (}  C^{e \mu}_{HL(1)} (M) 
+C^{e \mu}_{HL(3)} (M) {\Big )} 
 -0.0043C^{e\mu }_{e Z}(M) \ln \frac{M}{m_W}
 \nonumber \\
&&
- 59 C^{e \mu  tt}_{LEQU(3)}(M) \ln \frac{M}{m_W} 
- C^{e\mu  cc}_{LEQU(3)}(M)   \left( 0.43 \ln \frac{M}{m_W} + 1.5\right)
 \nonumber \\
&&+  0.039 C^{e \mu  tt}_{LEQU(1)}(M) \ln ^2  \frac{M}{m_W}   
+ 0.002 \left( 1 +  \ln \frac{M}{m_W}  \right)  C^{e \mu  cc}_{LEQU(1)}(M)
 \nonumber \\
&&- 4.8 \times 10^{-5} \ln^2 \frac{M}{m_W}
{\Big (}   C_{LU}^{e \mu tt }(M) +C_{LQ(1)}^{e \mu tt }(M) -3C_{LQ(3)}^{e \mu tt }(M) {\Big )}
\label{bornes2}
\eea
 gives the ``one-operator-at-a-time''
bounds listed in table
\ref{tab:bds}.  These  bounds are obtained
by assuming that one operator dominates
the $\meg$ amplitude, so  neglect interferences between 
the various coefficients. If both  the left-handed
dipole  $C_{e\g}^{ \mu e*}$ and the right-handed $C_{e\g}^{e \mu}$
are generated, then the right column  could be
divided by $\sqrt{2}$.   The  bounds of the first six rows agree
to within a factor 2 with the constraints given in \cite{PS},
who do not  constrain the coefficients given in 
the last four rows.  The vector operators, given
in the last two rows,  barely pass the
``constrainable'' threshhold defined above ($C <1 $ at $M = 100 m_t$).
This retroactively justifies that the mixing of vectors
into scalars was neglected, because
it  would be suppressed by an additional loop.

\begin{table}[htp!]
\begin{center}
\begin{tabular}{|ll|c|}
\hline
\hline
$ C^{\mu e *}_{e \g}$ &$ C^{e \mu }_{e \g}$&$1.2\times 10^{-8}$\\
\hline
$C^{\mu e *}_{e Z} \ln \frac{M}{m_W}$ 
&$ C^{e \mu }_{e Z} \ln \frac{M}{m_W}$ 
&$3.0\times 10^{-6}$\\
\hline
$ C^{\mu e tt*}_{LEQU(3)} \ln \frac{M}{m_W} $&
$ C^{e \mu tt}_{LEQU(3)} \ln \frac{M}{m_W} $  &
$2.0 \times 10^{-10}$\\
\hline
$ C^{\mu e cc*}_{LEQU(3)} (\ln \frac{M}{m_W} +3.5) $&
$ C^{e \mu cc}_{LEQU(3)} (\ln \frac{M}{m_W} +3.5) $  &
$2.8\times 10^{-8}$\\
\hline
$ C^{\mu e tt*}_{LEQU(1)} \ln^2 \frac{M}{m_W}  $&
$ C^{e \mu tt}_{LEQU(1)} \ln ^2 \frac{M}{m_W}  $&
$3.1\times 10^{-7}$\\
\hline
$ C^{\mu e cc*}_{LEQU(1)} ( \ln   \frac{M}{m_W}+1)  $&
$ C^{e \mu cc}_{LEQU(1)}( \ln  \frac{M}{m_W}+1)  $&
$6.0\times 10^{-6}$\\
\hline
$ C^{\mu e *}_{EH}$ &$ C^{e \mu }_{EH}$  &$7.5\times 10^{-7}$  \\
\hline
$C^{e \mu}_{HE}$ &$C^{e \mu }_{HL(1)}, C^{e \mu }_{HL(3)}$  &$1.2\times 10^{-5}$\\
\hline
$ C^{\mu e tt*}_{EQ} \ln^2 \frac{M}{m_W}  $&
$ C^{e \mu tt}_{LU} \ln ^2 \frac{M}{m_W}  $&
$2.5\times 10^{-4}$\\
\hline
$ C^{\mu e tt*}_{EU} \ln^2 \frac{M}{m_W}  $&
$ C^{e \mu tt}_{LQ(1)} \ln ^2 \frac{M}{m_W}  $,
$ 3 C^{e \mu tt}_{LQ(3)} \ln ^2 \frac{M}{m_W}  $
&
$2.5\times 10^{-4}$\\
\hline
\hline
\end{tabular}
\caption{ Approximate  ``one-operator-at-a-time'' 
constraints on operator coefficients  evaluated at  the scale $M$,
from the MEG bound \cite{MEG13}  on $BR(\meg)$, as given
in  eqns (\ref{bornes},\ref{bornes2}). For
a given choice of scale $M$, the quantity in either
left column should be less than  the number in
the right colomn  multiplied by ${M^2}/{m_t^2}$. The
operators are labelled in the same way as
the coefficients, and given in Appendix \ref{BWP}.
\label{tab:bds}}
\end{center}
\end{table}

\section{Discussion of the machinery and its application to $\meg$}
\label{sec:disc}


The MEG experiment \cite{MEG13} sets a stringent bound
on the dipole operator coefficients at  low energy (see
eqn (\ref{expt})). In translating this constraint to a scale
$M> m_W$, the analysis here aimed to  include
the ``Leading Order'' contribution of all ``constrainable''
operators, where LO was taken to mean numerically largest,
and an operator was deemed constrainable if a bound
$C<1$  could be obtained at $M\geq 100 m_t$.
{\color{black} However,   two-loop running, which gives the leading order
mixing of vectors to the dipole, was not included here,
so many constraints on vector operators are missing.} 
 As a result,
the  one-operator-at-a-time  limits given in table
\ref{tab:bds} are obtained from a combination of tree,
one- and two-loop matching, with RGEs at one-loop. 
Why do these multi-loop matching contributions  arise ?

First consider operator dimensions above
and below $m_W$. 
There is a rule of thumb  in EFT\cite{BurasHouches}, 
that one matches at a loop-order
lower than one runs, where  the loops are
counted in the interaction giving the running.
This makes sense if the loop
expansion is in one coupling, or if  the same diagram gives the
running and one-loop matching,  because the
running contribution is relatively enhanced
by the log. 
For instance, an electroweak box
diagram at $m_W$ generates a four-fermion
operator ``at tree level'' in QCD, which
can run down with 1-loop QCD RGEs.
One could hope that a similar  argument might apply
above $m_W$: a diagram giving one-loop
matching  could contribute to running
above $m_W$, so the subdominant matching
could be neglected. However, this
is not the case at $m_W$, because 
 SU(2)-invariant dimension-six operators 
from above $m_W$ can match onto operators that
would be dimension eight  if one imposed
SU(2), but that  are ${\cal O}(1/M^2)$
and dimension six in the QED$\times$QCD
invariant theory below $m_W$.
For example, the LFV $Z$ penguin operators 
given in eqns (\ref{LLpenguin}-\ref{EEpenguin})
 match at one -loop onto the
``dimension eight'' dipole 
$ y_\mu H^\dagger H(\overline{L}_e H \sigma\cdot F  E_\mu )$.
Similarly, the LFV Higgs interaction 
 $H^\dagger H(\overline{L}_e H  E_\mu)$ matches
at two-loop to the same ``would-be-dimension-eight'' dipole.
So the expectation that running dominates
matching can fail at $m_W$. 

The expectation that one loop is larger
than two-loop can fail when perturbing
in a hierarchy of Yukawa couplings. 
The dipole's affinity to Yukawas
arises because  the lepton chirality changes,
and the operator has a Higgs leg.  The dipole
operator here  is defined to include  a muon Yukawa
coupling $Y_\mu$ (see eqn (\ref{Lmeg})), because in many models,
the Higgs leg attaches to a Standard Model fermion, 
and/or  the lepton chirality
flips due to a Higgs coupling. And
while its difficult to avoid
the  $Y_\mu$ in one-loop contributions
to the dipole (see the discussion in 
\cite{meghdm}), there are more possibilities
at two-loop. In particular, 
it is ``well-known'' \cite{bjw}
that the leading contribution to $\meg$
of a flavour-changing Higgs interaction,
is via the two-loop top and $W$ diagrams
included in the matching contribution of
eqn (\ref{top}).

Its unclear to the author what to do about either of these problems.
Perhaps only the 
LFV operators with at least two Higgs legs
give their leading contributions in matching rather 
than running \footnote{In tree-level matching, the $Z$ penguins do give
their leading contribution to
four-fermion operators; its only  the ``leading contribution
to $\meg$'' which arises in one-loop matching. See 
the discussion in section
\ref{ssec:LO}.}.
And maybe performing  the matching and running at two-loop
would include the leading contributions in
loops, logs and Yukawa hierarchies. However,
a complete two-loop analysis would  take
some effort --- perhaps  it would be  simpler to
list all  the  possible  operators  at the scale $M$,
locate   their ``Leading Order'' contributions,
and include them.



  As discussed above, it is important to match with care at $m_W$.
A slightly different question
is  whether  its important to match onto  
the extended (non-SU(2)-invariant) operator basis at $m_W$?
The answer probably depends on the low energy
observables of interest. 
In the analysis here  of $\meg$,
 the four-fermion operators that were added  below $m_W$
(such  as the scalar four-fermion operators
${\cal O}^{e\mu bb}_{S,YY}$,
${\cal O}^{e\mu \tau \tau}_{S,YY}$ and
${\cal O}^{e\mu \mu \mu}_{S,YY}$
 given in eqn (\ref{opstauW})),
 are 
numerically irrelevant
{\it provided} that  the matching is performed
at two-loop. 
This is because they were generated in tree-matching
by the Higgs LFV operator $H^\dagger H(\overline{L} H  E)$,
suppressed by the $b,\tau$ or $\mu$ Yukawa coupling,
see eqns (\ref{24},\ref{27},\ref{40},\ref{42}).
Then, in  QED running, they
 mix to the dipole (possibly via the tensor),
which brings in another factor of the light fermion mass.
With tree matching, this is the best constraint on the Higgs LFV operator, so  is interesting to include.
However, it is irrelevant  compared with the two-loop
diagrams involving a top and $W$ loop,
which match the Higgs LFV operator directly
onto the dipole. This two-loop matching contribution
is relatively enhanced by a factor $\sim 100$
 as can be seen by comparing  the 
square brackets of eqns (\ref{square}) and (\ref{top}).
So in the case of $\meg$, it seems that
one would get the correct constraints on
operator coefficients at $M$ by using an
SU(2)-invariant four-fermion operator basis all the way 
between $m_\mu$ and $M$, provided the
matching at $m_W$  is performed to
whatever loop order retains the
``leading'' contributions. 


The QED mixing between $m_\mu$ and  $m_W$ 
modifies significantly the combination of operators
that are constrained by $\meg$. This is illustrated
in figure \ref{fig:c11}, which shows that the constraint
has rotated in operator space, to constrain
the linear combination of coefficients given in
eqn (\ref{plot}).  Coefficients of tensor operators that were  of
a similiar magnitude  to the dipole coefficient
could give significant enhancement or cancellations.
So the QED running is important.  In addition,
the MEG constraint on $BR(\meg)$ is restrictive ---
as discussed in section \ref{ssec:param}, it could
constrain  New Physics which contributes at one loop
up to a scale $M \sim 10^7$ GeV. So
it would be sensitive to two-loop contributions
from LFV operators at a scale of $10^5$ GeV. 
However,   
in matching at $m_W$ onto SU(2)-invariant dimension-six
operators, 
many of the tensor and scalar operators
which mix with the dipole below $m_W$, are generated
with small coefficients which give a negligeable contribution
to $\meg$. The point is that the scalars
and tensors involving leptons and $d$-type quarks
are generated by the Higgs LFV operator,
whose leading contribution to $\meg$ arises
in two-loop matching.


There are many improvements that could be made to these
estimates. 
 Including the experimental constraints
from $\meee$ and $\muc$ would  directly constrain the vector
operators, and give independent constraints on some of
the operators that contribute to $\meg$.  There are
more operators than constraints, so this could allow to
identify linear combinations of operators that
are not constrained.
 One-loop
matching is motivated by the
restrictive experimental bounds, which
allow to probe multi-loop effects. 
In addition, there are  operators which require one-loop
matching, such as
the two-gluon operators  relevant
to $\muc$. 
  Two-loop running  is required to
get the leading order contribution
of vector operators to $\meg$, and  could be interesting
above $m_W$ if  there are diagrams 
that dominate the  one-loop running
due to the presence of large Yukawas,
or if 
 quark flavour-off-diagonal  operators are included,
which may contribute to $\meg$ at two-loop
\cite{topLFV}. 
It is also motivated by the experimental sensitivity.
 Finally,  dimension eight operators
can be relevant if the New Physics
scale is not to high \cite{meghdm}.

\section{Summary}

This paper assumes that there is
new lepton flavour violating (LFV) physics
at a scale $M \gg m_W$, and
no  relevant other  new physics below. So  at scales below $M$,
LFV can be described in an Effective Field Theory
constructed with Standard Model fields
and dimension six operators.  The aim was to translate
experimental constraints on selected  $\mu \leftrightarrow e$
flavour changing processes, from the
low energy scale of the experiments to 
operator coefficients at the
scale $M$. As a first step,  this
paper reviews and compiles some of  the formalism
required to get from low energy to 
the weak scale: a QED$\times$QCD
invariant operator basis is given in section
\ref{sec:low}, the one-loop RGEs to run the coefficients
to $m_W$ are discussed in section \ref{sec:runQED},
 the anomalous dimensions mixing scalars, tensors
and dipoles are given in appendix \ref{sec:anomdim},
and tree matching  onto SU(2)-invariant
operators  at $m_W$  is presented in section \ref{@mW}.

As a simple application of  the formalism, the
experimental bounds on $\meg$ were translated to 
the scale $M$ in section \ref{sec:meg}. The process $\meg$
was chosen because it is an electromagnetic decay, and
constrains only the coefficients
of the two dipole operators.  The resulting
constraints  at $M$ on two linear combinations of
operators are given in eqn (\ref{bornes},\ref{bornes2}).
{\color{black}
These limits are approximative, due to the
many simplifications discussed in the paper, 
valid at best to one significant figure.}
Bounds on individual operators can be
obtained by assuming  one operator  dominates the
sum; the resulting  constraints  are listed in
table \ref{tab:bds}. At a scale $M\sim 100 m_t$,
$\meg$ is sensitive to over a dozen operators,
whereas, if $M \gsim 10^{7}$ GeV, then $\meg$  is sensitive
to only a few. 

The formalism  of the first sections did not work
well for $\meg$. Tree matching and one-loop
running missed the largest contributions of
some operators, as discussed in section \ref{sec:disc}.
This curious problem  could benefit
from more study, in order to identify a practical and systematic solution.

\subsection*{Acknowledgements}

I  am very grateful to  Junji Hisano  
for  interesting questions and discussions,
and thank Peter Richardson, Gavin Salam
and Aneesh Manohar for useful comments.

\appendix

\section{Operator normalisation}
\label{app:2}

All the operators  introduced section \ref{sec:low}
appear in the Lagrangian with a 
coefficient $-C/M^2$, and the operator normalisation is chosen
to ensure that the Feynman rule is $-i C/M^2$. This implies
a judicious distribution of $\frac{1}{2}$s, which
is the subject of this Appendix.

The $  {\cal O}^{e \mu}$ are flavour-changing, so can be imagined as
 off-diagonal elements of the matrix $  {\cal O}$ in lepton flavour space.  
They annihilate a $\mu$, and create an $e$, so 
 the hermitian
conjugate of the operator 
should appear
in the Lagrangian too. However, the Lagrangian is
a flavour-scalar, so in the Lagrangian is -$\frac{1}{M^2}$Tr$[ C {\cal O}]$,
where  the coefficients $C$  are also a matrix in flavour space.
(For instance, to obtain only  $  {\cal O}^{e \mu}$ in the Lagrangian,
one takes only  $ C^{e \mu} \neq 0$.)
Adding $+h.c.$ means adding  -$\frac{1}{M^2}$Tr$[  {\cal O}^\dagger C^\dagger]$.  
If  $ {\cal O}^\dagger =  {\cal O}$, as in the case of 
vector operators, then there are two possibilities
for the matrix-in-flavour-space $C$: either take  $C^\dagger = C$ 
(so if  $ C^{e \mu} \neq 0$,
then $C^{ \mu e }= C^{e \mu *}$), so
  Tr$[  {\cal O}^\dagger C^\dagger] = $ Tr$[ C {\cal O}]$.
Then in  the Lagrangian appears
 Tr$[  {\cal O}^\dagger C^\dagger] + $ Tr$[ C {\cal O}]$, 
so the operator  should be normalised with 1/2 
to compensate for  this double-counting,
and thereby ensure that the F-rule is $-i C^{e \mu}/M^2$.
Alternatively,  one does not impose $C^\dagger = C$,
and only puts the desired  $ C^{e \mu} \neq 0$ coupling
in the Lagrangian, where the $+h.c.$
generates the the anti-particle  amplitude, and the
Feynman rule is again  $-i  C^{e \mu}/M^2$, without the
factor of 1/2 in the operator definition.
Scalars and tensor operators   are not hermitian,
{\it eg}: $$
[S] = \left[ \begin{array}{cc}
\overline{e}  P_Y e & \overline{e}  P_Y \mu \\
\overline{\mu}  P_Y e  & \overline{\mu}  P_Y \mu \\
\end{array}
\right] ~~~,~~~
[S]^\dagger = \left[ \begin{array}{cc}
\overline{e}  P_X e & \overline{e}  P_X \mu    \\
\overline{\mu}  P_X e & \overline{\mu}  P_X \mu \\
\end{array}
\right] ~~~~X \neq Y
$$
so a  scalar or tensor operator $ {\cal O}^{e\mu}$
will induce two distinct $\mu \to e$ flavour-changing
interactions of different chirality.
In the case of the dipole, 
$[{\cal O}_{D,R }]^\dagger =[{\cal O}_{D,L }]$, so if
one writes
\beq
  - \frac{C^{e\mu}_{D,R } }{M^2}{\cal O}^{e\mu}_{D,R } 
 - \frac{C^{\mu e}_{D,R } }{M^2}{\cal O}^{\mu e}_{D,R } 
 - \frac{ C^{e\mu}_{D,L } }{M^2}{\cal O}^{e\mu}_{D,L }  
- \frac{  C^{\mu e}_{D,L } }{M^2}{\cal O}^{\mu e}_{D,L }
+ h.c.
\label{dipoles}
\eeq
then the $+h.c.$ is double-counting, it just adds
all the same operators a second time {\color{black}
(which implies $ C^{e\mu *}_{D,L } =  C^{\mu e}_{D,R }$,
 $ C^{e\mu *}_{D,R } =  C^{\mu e}_{D,L }$).
So  I   include  in ${\cal L}$  the first and third operators
of eqn (\ref{dipoles}),  and the $+h.c.$.}

\section{Anomalous dimension matrix in QED}
\label{sec:anomdim}

In this appendix are given the various sub-matrices of the
anomalous dimension matrix $\bm{\Gamma}_{STD}$  of equation 
(\ref{GammaSTD}). The relevant diagrams are given in 
figure \ref{fig:diag1}.

\ben

\item For scalar operators, the penguin diagrams (first and second)
do not contribute to one-loop mixing
among four-fermion operators, because the photon couples to the vector current.
However,  the second  penguin diagram, with on-shell photon (no
fermions)  mixes the ${\cal O}^{e\mu ll}_{S,YY}$ operators
for $l \in \{e,\mu\}$, to the dipole.
This gives a matrix  :
\beq
 \gamma_{S,D}^{l,} =
\begin{array}{c|cc}
&C_{D,L}^{e \mu} & C_{D,R}^{e \mu }
\\ \hline
C_{S,LL}^{e \mu ll}& {\color{black}-}\frac{m_l }{e m_\mu} &0\\
C_{S,RR}^{e \mu ll}&0& {\color{black}-} \frac{m_l }{e m_\mu} \\
\end{array}
\eeq

Diagrams 3 and 4 are the same as the 
 mass renormalisation diagrams ($\g_m = 6$ in QED), so combined with
the wave-function diagrams, they renormalise scalar operators, giving
a diagonal matrix:
\beq
\gamma_{S,S}^{f_1,f_2} =
\begin{array}{c|cccc}
&C_{S,LL}^{e \mu f f} & C_{S, RR}^{e \mu f f}& C_{S,LR}^{e \mu f f} & C_{S,RL}^{e \mu f f}\\ \hline
C_{S,LL}^{e \mu f f}& 6(1 + Q_f^2) &0&
0 &0\\
C_{S,RR}^{e \mu f f}&0& 6(1 + Q_f^2) &0&
0\\
C_{S,LR}^{e \mu f f}&0& 0& 6(1 + Q_f^2) &0\\
C_{S,RL}^{e \mu f f}&0&0&0& 6(1 + Q_f^2) \\
\end{array}
\eeq 
where the $(1 + Q_f^2)$ arises
from the photon  exchange across either current.

The last four  diagrams  mix the $YY$ scalars to the tensors
(the $YX$ tensor vanishes) with $\g = 2Q_f$:
\beq
\gamma_{S,T}^{f,f} = 
\begin{array}{c|cc}
&C_{T,LL}^{e \mu f f} & C_{T, RR}^{e \mu f f} \\ \hline
C_{S,LL}^{e \mu f f}& 2 Q_f &0 \\
C_{S,RR}^{e \mu f f}&0& 2Q_f \\
C_{S,LR}^{e \mu f f}&0& 0 \\
C_{S,RL}^{e \mu f f}&0&0\\
\end{array}
\eeq

\item 
 The tensors mix to the dipoles, via the first
diagram with the $f_2$ line removed.  This gives
\beq
\g_{T,D}^{f,~} =
\begin{array}{c|cc}
&C^{D,L}_{f} & C^{D, R}_{f} \\ \hline
C^{T,LL}_{f}&  8 \frac{Q_fN_c m_f}{ m_\mu e} &0 \\
C^{T,RR}_{f}&0& 8 \frac{Q_fN_c m_f}{ m_\mu e}  \\
\end{array}
\eeq
 The third and fourth diagrams do not
renormalise the tensors because $\g^\a \sigma \g_\a = 0$, 
but  the wavefunction diagrams do: 
\beq 
 \g_{T,T}^{f,f} =
\begin{array}{c|cc}
&C_{T,LL}^{e \mu f f} & C_{T, RR}^{e \mu f f} \\ \hline
C_{T,LL}^{e \mu f f}& -2(1 + Q_f^2) &0 \\
C_{T,RR}^{e \mu f f}&0&- 2(1 + Q_f^2)  \\
\end{array}
\eeq
and finally,  the last four diagrams  mix the tensors to scalars,
giving 
\beq \g_{T,S}^{f,f} =
\begin{array}{c|cc}
&C_{S,LL}^{e \mu f f} & C_{S, RR}^{e \mu f f} \\ \hline
C_{T,LL}^{e \mu f f}& -96 Q_f &0 \\
C_{T,RR}^{e \mu f f}&0& -96 Q_f  \\
\end{array}
\eeq
These tensor$\to$scalar mixing elements   of the
QED anomalous dimension matrix are large, suggesting
that one could redefine the operator basis to use 
a linear combination of scalar and tensor operators
with smaller off-diagonal elements. 
However,  QCD does not mix the scalars
and tensors, which favours them as basis operators.
In addition, the  tensor$\to$scalar mixing does not
enter the $\meg$ example of section \ref{sec:meg},
where the scalar-tensor operator basis  gives
the correct behaviour, as verified by
comparing  EFT and exact calculations of
$\meg$ in the 2HDM \cite{meghdm}.

{\color{black}
The dipole also renormalises itself \cite{CzarJ}, although
this effect is not included here:
\beq \g_{D,D} =
\left[
\begin{array}{cc}
 16  &0 \\
0& 16  \\
\end{array}
\right]
\eeq}

\item  The diboson  operators ${\cal O}_{GG,Y},{\cal O}_{FF,Y}$ are of
dimension 7, so the four-fermion operators and dipole operators
do not mix into them.

\een

\section{ Spinor Stuff}


The  Fiertz identities can be written  for chiral fermions as: 
\bea
 (\overline{a} P_L b)(\overline{c} P_R d)& =& -\frac{1}{2}(\overline{a} \gamma^{\mu} P_R d)
   (\overline{c} \gamma_{\mu} P_L b)
\label{F1} \\
 (\overline{a} \gamma^{\mu} P_{L,R} b)(\overline{c} \gamma_{\mu} P_{L,R} d)
 & =& (\overline{a} \gamma^{\mu} P_{L,R} d)(\overline{c} \gamma_{\mu} P_{L,R} b)
\label{F2} \\
 (\overline{a} P_X b)(\overline{c} P_Xd)& =& 
-\frac{1}{2}(\overline{a} P_X d)(\overline{c} P_X b)
-\frac{1}{8}(\overline{a} \sigma^{\nu \mu} P_X d)
(\overline{c} \sigma_{\nu \mu} P_X b)
\label{F3} \\
 (\overline{a}  \sigma^{\nu \mu} P_{X} b)(\overline{c}  \sigma_{\nu \mu} P_{X} d)
&  =& \frac{1}{2}
 (\overline{a}  \sigma^{\nu \mu} P_{X} d)(\overline{c}  \sigma_{\nu \mu} P_{X} b)
-6(\overline{a} P_X d)(\overline{c} P_X b)
\label{F4}
\eea
where the  relation 
$\sigma_{\mu \nu} = \frac{i}{2} 
  \varepsilon_{\mu \nu \a \b} \sigma^{\a \b} \g_5$,
was used to replace $\sigma$ with $\sigma P_X$.
It implies that
$(\overline{e}\sigma^{\a \b} \g_5 \mu) (\overline{\psi}\sigma_{\a \b} \g_5 \chi ) 
=
 (\overline{e} \sigma_{\mu \nu} 
\mu) (\overline{\psi}\sigma^{\mu \nu}  \chi ) $, so
\beq
(\overline{e}\sigma^{\a \b} P_Y \mu) (\overline{\psi}\sigma_{\a \b} P_Y \chi ) 
= \frac{1}{2}
(\overline{e}\sigma^{\a \b}  \mu) (\overline{\psi}\sigma_{\a \b}  \chi ) 
~~~,~~~
(\overline{e}\sigma^{\a \b} P_Y \mu) (\overline{\psi}\sigma_{\a \b} P_X \chi ) 
=0 ~~~(X\neq Y)
\label{vanish}
\eeq

\section{SU(2) invariant dimension six operators}
\label{BWP}

This Appendix lists dimension-six,  
SM-gauge invariant operators 
involving $e-\mu$ flavour change. 
The  operators are in the  Buchmuller-Wyler
basis, as pruned  in 
Grzadkowski {\it et.al.} \cite{polonais},
and this list is refered to as the BWP basis. The operators
are assumed to be added to the Lagrangian $+ h.c.$;
when this gives the $\bar{\mu} e$ operator,
it is not listed.  The $\tau^a$ are the Pauli matrices, with
$$\tau^2 = \left[ \begin{array}{cc} 0 &-i\\i & 0 
\end{array}\right]~~~.$$

The four-fermion operators involving 
$e$-$\mu$ flavour change and two quarks are:
\bea
{\cal O}_{L Q}^{(1) e \mu nm} &=&  \frac{1}{2}(\overline{L}_e \gamma^\a L_\mu ) 
(\overline{Q}_n \gamma^\a Q_m ) \label{OLQ1} \\
&=&  \frac{1}{2}[(\overline{e}_e \gamma^\a P_L  \mu )
+ (\overline{\nu}_e \gamma^\a P_L  \nu_\mu )]
[(\overline{u}_n \gamma^\a  P_L u_m ) 
+ (\overline{d}_n \gamma^\a  P_L d_m )]
\nonumber\\
{\cal O}_{L Q}^{(3)e \mu nm} &=&  \frac{1}{2}(\overline{L}_e \gamma^\a \tau^aL_\mu ) 
(\overline{Q}_n \gamma^\a \tau^a Q_m ) \label{OLQ3} \\
&=&   (\overline{\nu}_e \gamma^\a P_L  \mu )
(\overline{d}_n \gamma^\a  P_L u_m) 
+  (\overline{e}_e \gamma^\a P_L  \nu_\mu )
(\overline{u}_n \gamma^\a  P_L d_m ) \nonumber\\ 
&&~~~~~+  \frac{1}{2}[(\overline{\nu}_e \gamma^\a P_L  \nu_\mu ) -
(\overline{e}_e \gamma^\a P_L  \mu )]
[(\overline{u}_n \gamma^\a  P_L u_m ) 
- (\overline{d}_n \gamma^\a  P_L d_m )]
 \nonumber\\
{\cal O}^{e \mu nm}_{EQ} &=&  \frac{1}{2}(\overline{E}_e \gamma^\a E_\mu ) 
(\overline{Q}_n \gamma^\a Q_m ) \label{OEQ} \\
{\cal O}^{e \mu nm}_{LU} &=& \frac{1}{2} (\overline{L}_e \gamma^\a L_\mu ) 
(\overline{U}_n \gamma^\a U_m ) 
~~~~~~~~~~~~~~~~~~~~~~~~ \label{OLU}\\
{\cal O}^{e \mu nm}_{LD} &=& \frac{1}{2} (\overline{L}_e \gamma^\a L_\mu ) 
(\overline{D}_n \gamma^\a D_m ) 
~~~~~~~~~~~~~~~~~~~~~~~~ \label{OLD}\\
{\cal O}^{e \mu nm}_{EU} &=& \frac{1}{2} (\overline{E}_e \gamma^\a E_\mu ) 
(\overline{U}_n \gamma^\a U_m )
~~~~~~~~~~~~~~~~~~~~~~~~ \label{OEU}\\
{\cal O}^{e \mu nm}_{ED} &=& \frac{1}{2} (\overline{E}_e \gamma^\a E_\mu ) 
(\overline{D}_n \gamma^\a D_m )
~~~~~~~~~~~~~~~~~~~~~~~~ \label{OED}\\
{\cal O}^{e \mu nm}_{LE Q U}&=&  (\overline{L}_e^A  E_\mu )\epsilon_{AB} 
(\overline{Q}^B_n U_m )
~~~~~~~~~~~~~~~~~~~~~~~~
\label{scalaremu}\\
 &=& -(\overline{\nu}_e P_R \mu ) 
(\overline{d}_n P_R u_m ) + (\overline{e}_e P_R \mu ) 
(\overline{u}_nP_R u_m ) \nonumber \\
{\cal O}^{ \mu e nm}_{LE Q U}&=& (\overline{L}_\mu^A  E_e )\epsilon_{AB} 
(\overline{Q}^B_n U_m ) 
\label{scalarmue}\\
{\cal O}^{e \mu nm}_{LE  D Q}&=& (\overline{L}_e  E_\mu ) 
(\overline{D}_n Q_m )
~~~~~~~~~~~~~~~~~~~~~~~~
\label{scalarDemu}\\
 &=& (\overline{\nu}_e P_R \mu ) 
(\overline{d}_n P_L u_m ) + (\overline{e}_e P_R \mu ) 
(\overline{d}_nP_L d_m ) \nonumber \\
{\cal O}^{\mu e nm}_{LE  DQ }&=& (\overline{L}_\mu  E_e ) 
(\overline{D}_n Q_m )
~~~~~~~~~~~~~~~~~~~~~~~~
\label{scalarDmue}\\
{\cal O}^{e \mu nm}_{T, LEQU} &=&  (\overline{L}_e^A \sigma^{\mu\nu} E_\mu )\epsilon_{AB} 
(\overline{Q}^B_n \sigma_{\mu\nu} U_m )
\label{tensoremu} \\
{\cal O}^{\mu e nm}_{T, LEQU} &=&  (\overline{L}_\mu^A \sigma^{\mu\nu} E_e )\epsilon_{AB} 
(\overline{Q}^B_n \sigma_{\mu\nu} U_m )
\label{tensormue}  
\eea
where $L,Q$ are doublets and $E,U$ are singlets
(lower case are Dirac spinors, 
  SU(2) components selected  with $P_{L,R}$),
$n,m$ are possibly equal  quark family indices, and
$A,B$ are SU(2) indices.  The doublet quarks
are in the $d,s,b$ mass eigenstate basis.
The operator names are as in \cite{polonais} 
with $\phi \to H$; the flavour indices are in
 superscript. 

The  operators 
involving $e-\mu$ flavour change, and leptons, are:
\bea
{\cal O}^{e \mu ii}_{L L} &=& \frac{1}{2}(\overline{L}_e \gamma^\a L_\mu ) 
(\overline{L}_i \gamma^\a L_i ) \nonumber \\
&=& \frac{1}{2}[(\overline{e}_e \gamma^\a P_L  \mu )
+ (\overline{\nu}_e \gamma^\a P_L  \nu_\mu )]
[(\overline{\nu}_i \gamma^\a \tau^a P_L \nu_i ) 
+ (\overline{e}_i \gamma^\a \tau^a P_L e_i )]
\label{OLL}\\
{\cal O}^{e \mu ii}_{L E} &=& \frac{1}{2}(\overline{L}_e \gamma^\a L_\mu ) 
(\overline{E}_i \gamma^\a E_i ) 
\label{EL} \\
{\cal O}^{ii e \mu}_{ LE } &=& \frac{1}{2}
(\overline{L}_i \gamma^\a L_i ) (\overline{E}_e \gamma^\a E_\mu )   
\label{ELemu}\\
{\cal O}^{e \mu ii}_{EE} &=& \frac{1}{2}(\overline{E}_e \gamma^\a E_\mu ) 
(\overline{E}_i \gamma^\a E_i )
~~~~~~~~~~~~~~~~~~~~~~~~\\
-\frac{1}{2}{\cal O}^{e \tau \tau \mu}_{LE }&=& (\overline{L}_e  E_\mu )
(\overline{E}_\tau L_\tau )
~~,~~~~-\frac{1}{2}{\cal O}^{\mu \tau \tau e}_{LE }= (\overline{L}_\mu  E_e )
(\overline{E}_\tau L_\tau )
\label{scalarL}\\ &=& (\overline{\nu}_e P_R \mu ) 
(\overline{\tau} P_L \nu_\tau ) + (\overline{e}_e P_R \mu ) 
(\overline{\tau} P_L \tau ) \nonumber 
\eea
where 
$i$ is now a charged lepton family index,
and hermitian operators  are defined with a factor 1/2,
to agree with the factor of 1/2 present below $m_W$
as discussed in section \ref{app:2}.

The operator (\ref{scalarL}) 
appears  in the BWP basis in
 its Fierz-transformed version,
corresponding to the operator name given above.
Since here,  the $e$-$\mu$ flavour change
below $m_W$  remains  inside
a spinor contraction, the version are used interchangeably.

Then there are the operators  allowing interactions
with gauge bosons and Higgses.  This includes the dipoles,
which are normalised with  the muon Yukawa coupling so
as to match onto the normalisation of Kuno-Okada \cite{KO}:
\bea
{\cal O}^{e \mu}_{EH } =  H^\dagger H \overline{L}_e H  E_\mu 
&&
{\cal O}^{\mu e}_{E H } =  H^\dagger H \overline{L}_\mu H  E_e 
 \label{yuk6L} \\
{\cal O}^{e \mu}_{eW } =  y_\mu (\overline{L}_e  \vec{\tau}^a H \sigma^{\a \b} E_\mu ) W^a_{\a \b} &&
{\cal O}^{ \mu e}_{eW } =  y_\mu (\overline{L}_\mu   \vec{\tau}^a 
H \sigma^{\a \b}  E_e ) W^a_{\a \b}
 \label{magmo2L} \\
{\cal O}^{e \mu}_{eB } =   y_\mu(\overline{L}_e H \sigma^{\a \b}  E_\mu ) B_{\a \b}
&&
{\cal O}^{\mu e}_{eB } =  y_\mu (\overline{L}_\mu H \sigma^{\a \b}  E_e ) B_{\a \b}
 \label{magmo1L} \\%
{\cal O}^{(1) e\mu}_{HL } = i(\overline{L}_e \gamma^\a L_\mu ) 
(H^\dagger \stackrel{ \leftrightarrow}{  D_\a } H) &&
\label{LLpenguin} \\
{\cal O}^{(3)e\mu}_{HL } = i(\overline{L}_e \gamma^\a \vec{\tau}L_\mu ) 
(H^\dagger \stackrel{ \leftrightarrow}{  D_\a } \vec{\tau} H) && 
\label{LL3penguin} \\
{\cal O}^{e\mu}_{HE } = i(\overline{E}_e \gamma^\a E_\mu ) 
(H^\dagger \stackrel{ \leftrightarrow}{  D_\a } H) &&
\label{EEpenguin} 
\eea
where $i(H^\dagger \stackrel{ \leftrightarrow}{  D_\a } H) \equiv 
i(H^\dagger   D_\a H) - i( D_\a H)^\dagger H$, and $D_\a = \partial_\a + i \frac{g}{2}W_\a^a \tau^a + i \frac{g'}{2}{\color{black} Y}B_\a$. 
The sign in the 
covariant derivative fixes the sign of
the penguin operator and the SM Z vertex. 
{\color{black} These signs cancel in matching at $m_W$,
so the results of section 4 should be convention-independent.

This covariant derivative leads to
$D_\a = \partial_\a + i e Q A_\a$ after electroweak
symmetry breaking, giving a Feynman rule for the
photon-electron-electron vertex $i e \g^\m$. This
choice (opposite to Peskin-Schroeder but agrees with
Buras\cite{BurasHouches}),  controls
the sign of the  QED anomalous dimensions mixing four-fermion
operators to the dipole.  
}


\begin{thebibliography}{222222}

\bibitem{KO}
  Y.~Kuno and Y.~Okada,
  ``Muon decay and physics beyond the standard model,''
  Rev.\ Mod.\ Phys.\  {\bf 73} (2001) 151
  [hep-ph/9909265].

\bibitem{composite}
 F.~Feruglio, P.~Paradisi and A.~Pattori,
  ``Lepton Flavour Violation in Composite Higgs Models,''
  Eur.\ Phys.\ J.\ C {\bf 75} (2015) 12,  579
  doi:10.1140/epjc/s10052-015-3807-9
  [arXiv:1509.03241 [hep-ph]].

\bibitem{SUSY}

 J.~Hisano, T.~Moroi, K.~Tobe and M.~Yamaguchi,
  ``Exact event rates of lepton flavor violating processes in supersymmetric SU(5) model,''
  Phys.\ Lett.\ B {\bf 391} (1997) 341
   [Phys.\ Lett.\ B {\bf 397} (1997) 357]
  doi:10.1016/S0370-2693(96)01473-6
  [hep-ph/9605296].


J.~Hisano, D.~Nomura and T.~Yanagida,
  ``Atmospheric neutrino oscillation and large lepton flavor violation in the SUSY SU(5) GUT,''
  Phys.\ Lett.\ B {\bf 437} (1998) 351
  doi:10.1016/S0370-2693(98)00929-0
  [hep-ph/9711348].

  J.~Sato and K.~Tobe,
  ``Neutrino masses and lepton flavor violation in supersymmetric models with lopsided Froggatt-Nielsen charges,''
  Phys.\ Rev.\ D {\bf 63} (2001) 116010
  doi:10.1103/PhysRevD.63.116010
  [hep-ph/0012333].

{\color{black}
 A.~Ilakovac and A.~Pilaftsis,
  ``Flavor violating charged lepton decays in seesaw-type models,''
  Nucl.\ Phys.\ B {\bf 437} (1995) 491
  doi:10.1016/0550-3213(94)00567-X
  [hep-ph/9403398].

A.~Ilakovac and A.~Pilaftsis,
  ``Supersymmetric Lepton Flavour Violation in Low-Scale Seesaw Models,''
  Phys.\ Rev.\ D {\bf 80} (2009) 091902
  doi:10.1103/PhysRevD.80.091902
  [arXiv:0904.2381 [hep-ph]].

}


 A.~Masiero, P.~Paradisi and R.~Petronzio,
  ``Anatomy and Phenomenology of the Lepton Flavor Universality in SUSY Theories,''
  JHEP {\bf 0811} (2008) 042
  doi:10.1088/1126-6708/2008/11/042
  [arXiv:0807.4721 [hep-ph]].
 L.~Calibbi, P.~Paradisi and R.~Ziegler,
  ``Lepton Flavor Violation in Flavored Gauge Mediation,''
  Eur.\ Phys.\ J.\ C {\bf 74} (2014) 12,  3211
  doi:10.1140/epjc/s10052-014-3211-x
  [arXiv:1408.0754 [hep-ph]].
 P.~Paradisi,
  ``Constraints on SUSY lepton flavor violation by rare processes,''
  JHEP {\bf 0510} (2005) 006
  doi:10.1088/1126-6708/2005/10/006
  [hep-ph/0505046].
 A.~Abada, A.~J.~R.~Figueiredo, J.~C.~Romao and A.~M.~Teixeira,
  ``Probing the supersymmetric type III seesaw: LFV at low-energies and at the LHC,''
  JHEP {\bf 1108} (2011) 099
  doi:10.1007/JHEP08(2011)099
  [arXiv:1104.3962 [hep-ph]].

 F.~Feruglio, C.~Hagedorn, Y.~Lin and L.~Merlo,
  ``Lepton Flavour Violation in Models with A(4) Flavour Symmetry,''
  Nucl.\ Phys.\ B {\bf 809} (2009) 218
  doi:10.1016/j.nuclphysb.2008.10.002
  [arXiv:0807.3160 [hep-ph]].

 E.~Arganda and M.~J.~Herrero,
  ``Testing supersymmetry with lepton flavor violating tau and mu decays,''
  Phys.\ Rev.\ D {\bf 73} (2006) 055003
  doi:10.1103/PhysRevD.73.055003
  [hep-ph/0510405].

\bibitem{BR}
  A.~Brignole and A.~Rossi,
  ``Anatomy and phenomenology of mu-tau lepton flavor violation in the MSSM,''
  Nucl.\ Phys.\ B {\bf 701} (2004) 3
  doi:10.1016/j.nuclphysb.2004.08.037
  [hep-ph/0404211].


\bibitem{vector}
 R.~Kitano and K.~Yamamoto,
  ``Lepton flavor violation in the supersymmetric standard model with vector like leptons,''
  Phys.\ Rev.\ D {\bf 62} (2000) 073007
  doi:10.1103/PhysRevD.62.073007
  [hep-ph/0003063].

\bibitem{RS}
 R.~Kitano,
  ``Lepton flavor violation in the Randall-Sundrum model with bulk neutrinos,''
  Phys.\ Lett.\ B {\bf 481} (2000) 39
  doi:10.1016/S0370-2693(00)00444-5
  [hep-ph/0002279].

\bibitem{inverse}
 A.~Abada, D.~Das, A.~Vicente and C.~Weiland,
  ``Enhancing lepton flavour violation in the supersymmetric inverse seesaw beyond the dipole contribution,''
  JHEP {\bf 1209} (2012) 015
  doi:10.1007/JHEP09(2012)015
  [arXiv:1206.6497 [hep-ph]].
 F.~Deppisch and J.~W.~F.~Valle,
  ``Enhanced lepton flavor violation in the supersymmetric inverse seesaw model,''
  Phys.\ Rev.\ D {\bf 72} (2005) 036001
  doi:10.1103/PhysRevD.72.036001
  [hep-ph/0406040].

 E.~Arganda, M.~J.~Herrero, X.~Marcano and C.~Weiland,
  ``Imprints of massive inverse seesaw model neutrinos in lepton flavor violating Higgs boson decays,''
  Phys.\ Rev.\ D {\bf 91} (2015) 1,  015001
  doi:10.1103/PhysRevD.91.015001
  [arXiv:1405.4300 [hep-ph]].

\bibitem{seesaw}

 J.~Hisano, T.~Moroi, K.~Tobe and M.~Yamaguchi,
  ``Lepton flavor violation via right-handed neutrino Yukawa couplings in supersymmetric standard model,''
  Phys.\ Rev.\ D {\bf 53} (1996) 2442
  doi:10.1103/PhysRevD.53.2442
  [hep-ph/9510309].

 J.~Hisano and D.~Nomura,
  ``Solar and atmospheric neutrino oscillations and lepton flavor violation in supersymmetric models with the right-handed neutrinos,''
  Phys.\ Rev.\ D {\bf 59} (1999) 116005
  doi:10.1103/PhysRevD.59.116005
  [hep-ph/9810479].

 J.~Hisano and K.~Tobe,
  ``Neutrino masses, muon g-2, and lepton flavor violation in the supersymmetric seesaw model,''
  Phys.\ Lett.\ B {\bf 510} (2001) 197
  doi:10.1016/S0370-2693(01)00494-4
  [hep-ph/0102315].

A.~Abada, C.~Biggio, F.~Bonnet, M.~B.~Gavela and T.~Hambye,
  ``$\meg$ and $\tlg$ decays in the fermion triplet seesaw model,''
  Phys.\ Rev.\ D {\bf 78} (2008) 033007
  doi:10.1103/PhysRevD.78.033007
  [arXiv:0803.0481 [hep-ph]].



\bibitem{higgs}

 Y.~Omura, E.~Senaha and K.~Tobe,
  ``Lepton-flavor-violating Higgs decay $h \to \mu\tau$ and muon anomalous magnetic moment in a general two Higgs doublet model,''
  JHEP {\bf 1505} (2015) 028
  doi:10.1007/JHEP05(2015)028
  [arXiv:1502.07824 [hep-ph]].

A.~Abada and I.~Hidalgo,
  ``Neutrinos and lepton flavour violation in the left-right twin Higgs model,''
  Phys.\ Rev.\ D {\bf 77} (2008) 113013
  doi:10.1103/PhysRevD.77.113013
  [arXiv:0711.1238 [hep-ph]].
xM.~Blanke, A.~J.~Buras, B.~Duling, A.~Poschenrieder and C.~Tarantino,
  ``Charged Lepton Flavour Violation and (g-2)(mu) in the Littlest Higgs Model with T-Parity: A Clear Distinction from Supersymmetry,''
  JHEP {\bf 0705} (2007) 013
  doi:10.1088/1126-6708/2007/05/013
  [hep-ph/0702136].

  E.~Arganda, M.~J.~Herrero, X.~Marcano and C.~Weiland,
  ``Lepton flavour violating Higgs decays,''
  arXiv:1406.0384 [hep-ph].



\bibitem{4gen}
 A.~J.~Buras, B.~Duling, T.~Feldmann, T.~Heidsieck, C.~Promberger and S.~Recksiegel,
  ``Patterns of Flavour Violation in the Presence of a Fourth Generation of Quarks and Leptons,''
  JHEP {\bf 1009} (2010) 106
  doi:10.1007/JHEP09(2010)106
  [arXiv:1002.2126 [hep-ph]].


\bibitem{ZeeBabu}
J.~Herrero-Garcia, M.~Nebot, N.~Rius and A.~Santamaria,
  ``Testing the Zee-Babu model via neutrino data, lepton flavour violation and direct searches at the LHC,''
  arXiv:1410.2299 [hep-ph].


\bibitem{Georgi}
  H.~Georgi,
  ``Effective field theory,''
  Ann.\ Rev.\ Nucl.\ Part.\ Sci.\  {\bf 43 } (1993)  209-252.\\
  H.~Georgi,
  ``On-shell effective field theory,''
  Nucl.\ Phys.\  {\bf B361 } (1991)  339-350.
 

\bibitem{KKO}
  R.~Kitano, M.~Koike and Y.~Okada,
  ``Detailed calculation of lepton flavor violating muon electron conversion rate for various nuclei,''
  Phys.\ Rev.\ D {\bf 66} (2002) 096002
   [Phys.\ Rev.\ D {\bf 76} (2007) 059902]
  [hep-ph/0203110].

\bibitem{CKOT}
  V.~Cirigliano, R.~Kitano, Y.~Okada and P.~Tuzon,
  ``On the model discriminating power of mu ---> e conversion in nuclei,''
  Phys.\ Rev.\ D {\bf 80} (2009) 013002
  [arXiv:0904.0957 [hep-ph]].


\bibitem{CrivellinMuc}
  A.~Crivellin, M.~Hoferichter and M.~Procura,
  ``Improved predictions for $\mu\to e$ conversion in nuclei and Higgs-induced lepton flavor violation,''
  Phys.\ Rev.\ D {\bf 89} (2014) 093024
  [arXiv:1404.7134 [hep-ph]].

\bibitem{BW}
  W.~Buchmuller and D.~Wyler,
  ``Effective Lagrangian Analysis of New Interactions and Flavor Conservation,''
  Nucl.\ Phys.\ B {\bf 268} (1986) 621.
  doi:10.1016/0550-3213(86)90262-2





\bibitem{polonais}
B.~Grzadkowski, M.~Iskrzynski, M.~Misiak and J.~Rosiek,
  ``Dimension-Six Terms in the Standard Model Lagrangian,''
  JHEP {\bf 1010} (2010) 085
  [arXiv:1008.4884 [hep-ph]].

\bibitem{JMT}
  R.~Alonso, E.~E.~Jenkins, A.~V.~Manohar and M.~Trott,
  ``Renormalization Group Evolution of the Standard Model Dimension Six Operators III: Gauge Coupling Dependence and Phenomenology,''
  JHEP {\bf 1404} (2014) 159
  [arXiv:1312.2014 [hep-ph]].
 E.~E.~Jenkins, A.~V.~Manohar and M.~Trott,
  ``Renormalization Group Evolution of the Standard Model Dimension Six Operators II: Yukawa Dependence,''
  JHEP {\bf 1401} (2014) 035
  doi:10.1007/JHEP01(2014)035
  [arXiv:1310.4838 [hep-ph]].

\bibitem{CzarJ}
  A.~Czarnecki and E.~Jankowski,
  ``Electromagnetic suppression of the decay $\mu \to e \gamma$,''
  Phys.\ Rev.\ D {\bf 65} (2002) 113004
  doi:10.1103/PhysRevD.65.113004
  [hep-ph/0106237].

\bibitem{PepeGian}
  G.~Degrassi and G.~F.~Giudice,
  ``QED logarithms in the electroweak corrections to the muon anomalous magnetic moment,''
  Phys.\ Rev.\ D {\bf 58} (1998) 053007
  doi:10.1103/PhysRevD.58.053007
  [hep-ph/9803384].

\bibitem{PS}
  G.~M.~Pruna and A.~Signer,
  ``The $\mu\to e\gamma$ decay in a systematic effective field theory approach with dimension 6 operators,''
  JHEP {\bf 1410} (2014) 14
  [arXiv:1408.3565 [hep-ph]].


\bibitem{CNR}
  A.~Crivellin, S.~Najjari and J.~Rosiek,
  ``Lepton Flavor Violation in the Standard Model with general Dimension-Six Operators,''
  JHEP {\bf 1404} (2014) 167
  [arXiv:1312.0634 [hep-ph]].



\bibitem{GKM}
  T.~Goto, R.~Kitano and S.~Mori,
  ``Lepton flavor violating $Z$-boson couplings from nonstandard Higgs interactions,''
  Phys.\ Rev.\ D {\bf 92} (2015) 075021
  doi:10.1103/PhysRevD.92.075021
  [arXiv:1507.03234 [hep-ph]].

\bibitem{BBL}
G.~Buchalla, A.~J.~Buras and M.~E.~Lautenbacher,
  ``Weak decays beyond leading logarithms,''
  Rev.\ Mod.\ Phys.\  {\bf 68} (1996) 1125
  doi:10.1103/RevModPhys.68.1125
  [hep-ph/9512380].

\bibitem{BurasHouches}
  A.~J.~Buras,
  ``Weak Hamiltonian, CP violation and rare decays,''
  hep-ph/9806471.

\bibitem{Lusignoli}
  M.~Lusignoli,
  ``Electromagnetic Corrections to the Effective Hamiltonian for Strangeness Changing Decays and $\epsilon^\prime / \epsilon$,''
  Nucl.\ Phys.\ B {\bf 325} (1989) 33.
  doi:10.1016/0550-3213(89)90371-4

\bibitem{ACFG}
  J.~Aebischer, A.~Crivellin, M.~Fael and C.~Greub,
  ``Matching of gauge invariant dimension 6 operators for $b\to s$ and $b\to c$ transitions,''
  arXiv:1512.02830 [hep-ph].

\bibitem{SVZ78}
  M.~A.~Shifman, A.~I.~Vainshtein and V.~I.~Zakharov,
  ``Remarks on Higgs Boson Interactions with Nucleons,''
  Phys.\ Lett.\ B {\bf 78} (1978) 443.


\bibitem{HisanoDM}
  J.~Hisano, K.~Ishiwata and N.~Nagata,
  ``Gluon contribution to the dark matter direct detection,''
  Phys.\ Rev.\ D {\bf 82} (2010) 115007
  [arXiv:1007.2601 [hep-ph]].

\bibitem{Falkowski}
  A.~Falkowski and K.~Mimouni,
  ``Model independent constraints on four-lepton operators,''
  arXiv:1511.07434 [hep-ph].

\bibitem{ATLASLFV}
  G.~Aad {\it et al.} [ATLAS Collaboration],
  ``Search for the lepton flavor violating decay $\Zme$ in pp collisions at $\sqrt{s}=8$  TeV with the ATLAS detector,''
  Phys.\ Rev.\ D {\bf 90} (2014) 7,  072010
  doi:10.1103/PhysRevD.90.072010
  [arXiv:1408.5774 [hep-ex]].

\bibitem{DLV}
  S.~Davidson, S.~Lacroix and P.~Verdier,
  ``LHC sensitivity to lepton flavour violating Z boson decays,''
  JHEP {\bf 1209} (2012) 092
  doi:10.1007/JHEP09(2012)092
  [arXiv:1207.4894 [hep-ph]].

\bibitem{t3l}
  K.~Hayasaka {\it et al.},
  ``Search for Lepton Flavor Violating Tau Decays into Three Leptons with 719 Million Produced Tau+Tau- Pairs,''
  Phys.\ Lett.\ B {\bf 687} (2010) 139
  doi:10.1016/j.physletb.2010.03.037
  [arXiv:1001.3221 [hep-ex]].



\bibitem{DMEFT}
  S.~Davidson,
  ``Including the Z in an Effective Field Theory for dark matter at the LHC,''
  JHEP {\bf 1410} (2014) 84
  doi:10.1007/JHEP10(2014)084
  [arXiv:1403.5161 [hep-ph]].

\bibitem{meee}
  U.~Bellgardt {\it et al.} [SINDRUM Collaboration],
  ``Search for the Decay mu+ ---> e+ e+ e-,''
  Nucl.\ Phys.\ B {\bf 299} (1988) 1.
  doi:10.1016/0550-3213(88)90462-2

\bibitem{tlg}
  B.~Aubert {\it et al.} [BaBar Collaboration],
  ``Searches for Lepton Flavor Violation in the Decays $\tmg$
 and $\teg$,''
  Phys.\ Rev.\ Lett.\  {\bf 104} (2010) 021802
  doi:10.1103/PhysRevLett.104.021802
  [arXiv:0908.2381 [hep-ex]].

 K.~Hayasaka {\it et al.} [Belle Collaboration],
  ``New search for $\tmg$ and $\teg$ decays at Belle,''
  Phys.\ Lett.\ B {\bf 666} (2008) 16
  doi:10.1016/j.physletb.2008.06.056
  [arXiv:0705.0650 [hep-ex]].

\bibitem{AGC}
  R.~Alonso, B.~Grinstein and J.~Martin Camalich,
  ``$SU(2)\times U(1)$ gauge invariance and the shape of new physics in rare $B$ decays,''
  Phys.\ Rev.\ Lett.\  {\bf 113} (2014) 241802
  doi:10.1103/PhysRevLett.113.241802
  [arXiv:1407.7044 [hep-ph]].





\bibitem{MEG13}
  J.~Adam {\it et al.} [MEG Collaboration],
  ``New constraint on the existence of the $\mu^+ \to e^+\gamma$ decay,''
  Phys.\ Rev.\ Lett.\  {\bf 110} (2013) 201801
  [arXiv:1303.0754 [hep-ex]].

\bibitem{CHK}
  D.~Chang, W.~S.~Hou and W.~Y.~Keung,
  ``Two loop contributions of flavor changing neutral Higgs bosons to mu $\to$
  e gamma,''
  Phys.\ Rev.\  D {\bf 48} (1993) 217
  [arXiv:hep-ph/9302267].
  

\bibitem{DMPS}
  S.~Davidson, M.~L.~Mangano, S.~Perries and V.~Sordini,
  ``Lepton Flavour Violating top decays at the LHC,''
  Eur.\ Phys.\ J.\ C {\bf 75} (2015) 9,  450
  doi:10.1140/epjc/s10052-015-3649-5
  [arXiv:1507.07163 [hep-ph]].

\bibitem{meghdm}
 S.~Davidson,
  ``Mu to e gamma in the 2 Higgs Doublet Model: an exercise in EFT,''
  arXiv:1601.01949 [hep-ph].

\bibitem{bjw}
  J.~D.~Bjorken and S.~Weinberg,
  ``A Mechanism for Nonconservation of Muon Number,''
  Phys.\ Rev.\ Lett.\  {\bf 38} (1977) 622.
  doi:10.1103/PhysRevLett.38.622

\bibitem{topLFV}
 S.~Davidson, M.~L.~Mangano, S.~Perries and V.~Sordini,
  ``Lepton Flavour Violating top decays at the LHC,''
  Eur.\ Phys.\ J.\ C {\bf 75} (2015) 9,  450
  doi:10.1140/epjc/s10052-015-3649-5
  [arXiv:1507.07163 [hep-ph]].










\end{thebibliography}
\end{document}